\documentclass[12pt,a4paper]{article}
\usepackage{amsmath,amsthm,amssymb,geometry}
\usepackage{authblk,graphicx,float}
\usepackage{braket}
\usepackage[colorlinks=true,linkcolor=blue,
            citecolor=blue,urlcolor=blue,
            breaklinks=true]{hyperref}

\newtheorem{theorem}{Theorem}
\newtheorem{lemma}[theorem]{Lemma}
\newtheorem{corollary}[theorem]{Corollary}
\newtheorem{proposition}[theorem]{Proposition}   

\newtheorem{definition}{Definition} 
\newtheorem{notation}{Notation}

\date{}

\title{The logical structure of contextuality and nonclassicality}

\author[1]{Songyi Liu \thanks{liusongyi@buaa.edu.cn}}
\author[1,*]{Yongjun Wang \thanks{wangyj@buaa.edu.cn}}
\author[1]{Baoshan Wang \thanks{bwang@buaa.edu.cn}}
\author[1]{Chang He \thanks{hechang@buaa.edu.cn}}
\author[1]{Jincheng Wang \thanks{22091008@buaa.edu.cn}}
\affil[1]{School of Mathematical Sciences, Beihang University, Beijing, 100191, China}

\begin{document}

\maketitle

\begin{abstract}
Quantum contextuality represents a fundamental form of nonclassicality in quantum mechanics. To provide a more complete characterization of nonclassical properties in quantum systems, we adopt a logical perspective and propose a mathematical framework based on exclusive partial Boolean algebras (epBAs). This framework enables a unified description of contextuality and nonclassicality across finite general, quantum, and classical systems. We establish a unified and minimal classical counterpart for any finite general system. Within this framework, we formalize major categories of quantum contextuality, demonstrating that: 12 projectors suffice to generate Kochen-Specker scenarios; 10 projectors suffice to witness state-independent contextuality; and 3 observables suffice to witness quantum contextuality. Finally, we prove that contextuality is a sufficient but not necessary condition for nonclassicality.
\end{abstract}


\section*{Declarations}

\noindent\textbf{Competing interests.}
The authors have no relevant financial or non-financial interests to disclose.

\vspace{10pt}

\noindent\textbf{Acknowledgments.}
The work was supported by National Natural Science Foundation of China (Grant No. 12371016, 11871083) and National Key R\&D Program of China (Grant No. 2020YFE0204200).

\section{Introduction}
The establishment of quantum mechanics in the early 20th century spurred investigations into the mathematical foundations of nature. Some researchers posited that there exists hidden variables could simultaneously determine the values of all observables \cite{Fine1990Einstein}. The assumptions of \emph{realism} and \emph{locality} in the Einstein-Podolsky-Rosen (EPR) argument follow directly from the existence of hidden variables \cite{Einstein1935Can}. In the 1960s, Bell experiments \cite{Bell1964On,Clauser1969Proposed} and Kochen-Specker theorem \cite{Kochen1967The} demonstrated that quantum mechanics does not admit hidden variables, thereby invalidating the EPR argument. These results gave rise to the fields of Bell nonlocality \cite{Brunner2014Bell} and quantum contextuality \cite{Budroni2022Kochen}.

Contextuality characterizes violations of hidden-variable theories, with nonlocality representing a special case for space-like separated scenarios (Bell scenarios). Such violations observable within quantum mechanics are termed \emph{quantum contextuality}. This phenomenon has been experimentally verified in numerous setups, including the Clauser-Horne-Shimony-Holt (CHSH) experiment \cite{Clauser1969Proposed}, the Klyachko-Can-Binicioglu-Shumovsky (KCBS) experiment \cite{Alexander2008Simple, Lapkiewicz2011Experimental}, among others \cite{Kirchmair2009State, Amselem2009State, Guhne2010Compatibility, Moussa2010Testing, Qu2020Experimental}.

The contextuality of quantum mechanics is now firmly established. Quantum contextuality is regarded as a crucial resource within quantum information theory. However, a subsequent and more practical question arises: how can a \emph{finite quantum system} demonstrate contextuality? In applications that leverage contextuality for computation, communication, and cryptography \cite{Mark2014Contextuality, Gupta2023Quantum, Pirandola2020Advances}, one is constrained to finite quantum resources, namely, a finite set of quantum states and measurements. Notably, not all quantum systems exhibit nonclassical properties. For instance, an experiment capable of measuring only a single dichotomic observable (such as the spin of a spin-1/2 particle along the $x$-axis) is formally equivalent to tossing a classical coin. Similarly, the sequential measurement of two incompatible dichotomic observables can be simulated by tossing two classical coins in succession.

This consideration leads to the formal concept of a \emph{scenario} in contextuality theories. A scenario specifies the set of allowed measurements or observable events in an experiment. Let $X = \{A_1, \dots, A_n\}$ be a set of observables, and let $\mathcal{M}$ be a family of subsets of $X$. A subset $C\in\mathcal{M}$ is termed a \emph{context}, defined as a set of jointly commeasurable observables. Given a set $O$ of possible measurement outcomes, an assignment $e: C \to O$ (or $e \in O^C$) is called an \emph{event} on the context $C$. In frameworks such as the marginal problem \cite{Kurzynski2012Entropic, Fritz2013Entropic}, the family $\mathcal{M}$ of contexts defines the scenario. In the sheaf-theoretic approach \cite{Abramsky2011sheaf}, a scenario is a triple $(X, \frown, O)$, where $\frown$ denotes the compatibility relation on $X$. Graph-theoretic approaches \cite{Adan2014Graph, Acin2015A} define a scenario via the exclusivity graph or hypergraph of events. We refer to these definitions collectively as \emph{observable-based}, because events are defined by the outcomes of specific observables.

The scenario provides a framework for comparing quantum and classical systems, as the set $X$ can represent either quantum or classical observables. The probability distributions over measurement outcomes on a scenario are referred to by various names across different frameworks: models \cite{Fritz2013Entropic}, empirical models \cite{Abramsky2011sheaf}, probabilistic models \cite{Acin2015A}, behaviours \cite{Amaral2018On}, or states \cite{Abramsky2020The, Liu2025Atom}. In the classical case, these distributions are determined by hidden variables, called \emph{noncontextual}. In contrast, quantum mechanics and other generalized theories admit distributions that are unattainable classically, which are called \emph{contextual}. Within observable-based approaches, the existence of a contextual distribution in a quantum scenario provides definitive evidence against the existence of hidden variables.

While observable-based theories of contextuality have achieved considerable success, they encounter challenges when applied to scenarios defined by rank-1 projectors, such as the KCBS scenario \cite{Alexander2008Simple} and the Yu-Oh scenario \cite{Yu2012State}. The latter is recognized as the minimal vector set that exhibits state-independent contextuality \cite{Cabello2016Quantum}.

To illustrate the limitation, consider the following example. Define the vectors:
\begin{align*}
\ket{0}&= (0,0,1)^T, \quad \ket{1} = (0,1,0)^T, \quad \ket{2} = (1,0,0)^T, \\
\ket{x}&= (1/\sqrt{2},1/\sqrt{2},0)^T, \quad \ket{y} = (1/\sqrt{2},-1/\sqrt{2},0)^T,
\end{align*}
and two observables:
\begin{align*}
\hat{A} &= a_0 \ket{0}\!\bra{0} + a_1 \ket{1}\!\bra{1} + a_2 \ket{2}\!\bra{2}, \\
\hat{B} &= b_0 \ket{0}\!\bra{0} + b_1 \ket{x}\!\bra{x} + b_2 \ket{y}\!\bra{y}.
\end{align*}
The events $\hat{A}=a_0$ and $\hat{B}=b_1$ are exclusive. However, within the observable-based framework, the scenario is described the set of observables $X = \{A, B\}$ and the family of contexts $\mathcal{M} = \{\{A\}, \{B\}\}$, since $\hat{A}$ and $\hat{B}$ are incompatible. Consequently, the events $A=a_0$ and $B=b_1$ are not considered exclusive. This is because no jointly measurable observable exists that can simultaneously distinguish between them \cite{Adan2014Graph}. This example demonstrates that the observable-based definition fails to fully capture the structure inherent in a quantum scenario.

To provide a more complete characterization of scenarios, we introduce a perspective distinct from the observable-based approach, known as the \emph{event-based} or \emph{logical} framework. This perspective is rooted in quantum logic, a field pioneered by Birkhoff and von Neumann in the 1930s \cite{Birkhoff1936The}. Subsequent developments have yielded various formalisms, including partial Boolean algebras \cite{Kochen1967The}, propositional lattices \cite{Coecke2002Quantum}, Topos quantum logic \cite{Isham1998Topos, Doering2010Topos}, and frameworks based on observable algebras and context categories \cite{Frembs2023Gleason, Frembs2025Coming}.

In a nutshell, quantum logic investigates \emph{quantum events} and the logical relationships among them. Let $\hat{A}$ be a bounded Hermitian operator. The event $\hat{A}=a$ signifies that a measurement of $\hat{A}$ yields the outcome $a$. This is operationally equivalent to measuring the projector $\hat{P}_a$ and obtaining the outcome $1$, where $\hat{P}_a$ projects onto the eigenspace corresponding to the eigenvalue $a$ of $\hat{A}$. Thus, quantum events are fundamentally represented by projectors (whereas classical events correspond to subsets of a hidden variable space). Logical operations on events are then naturally defined through algebraic operations on projectors: the product $\hat{P}\cdot\hat{Q}$ corresponds to the conjunction $\hat{P} \land\hat{Q}$, and $\mathbf{I} - \hat{P}$ corresponds to the negation $\neg\hat{P}$.

Many frameworks presuppose that such logical operations are fully-defined, even for incompatible projectors $\hat{P}$ and $\hat{Q}$. However, in quantum mechanics, the conjunction of incompatible projectors lacks a direct and universally accepted physical interpretation \cite{Foulis2006Quantum, Kochen2015Reconstruction}. Therefore, we adopt the framework of \emph{partial Boolean algebras}, originally introduced by Kochen and Specker in their seminal proof that quantum mechanics admits no hidden variables \cite{Kochen1967The}. This framework only defines logical operations for compatible events, i.e.
\[e_1\land e_2\text{ is defined only if }e_1\odot e_2, \]
making it particularly well-suited for an analysis of experimentally observable events. In recent years, several works have successfully forged a connection between partial Boolean algebras and quantum contextuality \cite{Abramsky2020The, Liu2025Atom}, providing a natural foundation for our approach.

By defining logical operations solely for compatible projectors, the set $\mathbf{P}(\mathcal{H})$ of projectors on a Hilbert space $\mathcal{H}$ constitutes a partial Boolean algebra. The quantum scenario is then characterized by a subalgebra $\mathcal{Q}$ of $\mathbf{P}(\mathcal{H})$. A pivotal feature of $\mathcal{Q}$ is its closure under the defined logical operations. If a quantum experiment can observe an event $\hat{P} \in\mathcal{Q}$, then it can also observe the event $\neg\hat{P}$, signifying the non-occurrence of $\hat{P}$. Similarly, if $\hat{P}, \hat{Q} \in\mathcal{Q}$ are commeasurable, then the event $\hat{P} \land\hat{Q}$, representing the joint occurrence of $\hat{P}$ and $\hat{Q}$, is also observable. Given that $\{\neg, \land\}$ forms an adequate set of connectives, $\mathcal{Q}$ encompasses all observable events for a specific quantum experiment. Consequently, $\mathcal{Q}$ provides a complete characterization of the logical structure of a quantum scenario.

Given a quantum scenario $\mathcal{Q}$, a quantum state $\rho$ induces a probability distribution over it. The tuple $(\mathcal{Q}, \rho)$ thus fully describes a (static) quantum experiment, which we define as a \emph{quantum system}. Correspondingly, we define a \emph{classical system} as a tuple $(\mathcal{B}, p_{\mathcal{B}})$, where $\mathcal{B}$ is a Boolean algebra, representing the classical event space, and $p_{\mathcal{B}}$ is a state (probability distribution) on $\mathcal{B}$. Generally, a \emph{general system} is a tuple $(\mathcal{A},p)$, where $\mathcal{A}$ is an exclusive partial Boolean algebra (epBA) and $p$ is a state on $\mathcal{A}$.

To demonstrate the contextuality of a general system $(\mathcal{A}, p)$, one must show that it cannot be realized by any classical system. Moving beyond witnessing the violation of specific inequalities derived from hidden-variable theories, we establish a unified and minimal classical counterpart for any scenario $\mathcal{A}$: the power-set algebra (a Boolean algebra) of its deterministic states:
\[\mathcal{A}^c := \mathcal{P}(s_d(\mathcal{A})).\]
We will prove that $(\mathcal{A}, p)$ can be realized by a classical system if and only if it can be realized by $(\mathcal{A}^c, p_{\mathcal{A}^c})$ for some classical state $p_{\mathcal{A}^c}$. This result provides a unified framework for characterizing contextuality and nonclassicality of quantum systems, and we will demonstrate that contextuality is a sufficient but not necessary condition of nonclassicality.

We show that the logical structures of scenarios reveal a richer spectrum of nonclassical properties than those captured by observable-based theories. It enables the use of \emph{generating set} to characterize quantum scenarios, thereby reducing the number of projectors or observables required to witness nonclassicality and contextuality. For instance, the minimal Kochen-Specker set \cite{Cabello1997Bell,Xu2020Proof} and state-independent contextuality (SIC) set \cite{Yu2012State,Cabello2016Quantum} consists of 18 and 13 vectors, while they can be represented by 12 and 10 projectors. The minimum number of observables to witness quantum contextuality is considered as 4 in observable-based theories \cite{Kurzynski2012Entropic, Xu2019Necessary}, while this number can be reduced to 3 within the logical framework.

The remainder of this paper is organized as follows. In Section~\ref{sec-pBA}, we introduce the framework of partial Boolean algebra. Section~\ref{sec-CS and QS} provides a characterization of classical and quantum systems. In Section~\ref{sec-classical embedding}, we present the classical embedding to characterize contextuality and nonclassicality, proving that it provides a unified and minimal classical counterpart for any finite general scenario. Section~\ref{sec-QNC} formalizes the major categories of quantum contextuality, including Kochen-Specker scenarios, state-independent contextuality (SIC), and state-dependent contextuality (SDC). Building upon well-known contextual examples, we obtain several novel results using the tool of (perfect) generating sets: 12 projectors generate Kochen-Specker scenarios (Subsection~\ref{subsec-KS}); 10 projectors witness SIC (Subsection~\ref{subsec-SIC}) and 3 observables witness SDC (Subsection~\ref{subsec-SDC}). We also present an example showing that contextuality is not necessary for nonclassicality (Subsection~\ref{subset-NC without C}). Section~\ref{sec-conclusion} presents the conclusion and outlook.

\section{Partial Boolean algebra}\label{sec-pBA}

This section introduces the fundamental concepts of partial Boolean algebras. For comprehensive treatments, we refer the reader to \cite{Kochen1967The, Van2012Noncommutativity, Abramsky2020The, Liu2025Atom}.

\vspace{10pt}
\begin{definition}
    A \textbf{partial Boolean algebra} (\rm{pBA}) is a structure $(\mathcal{A},\odot,\lnot,\land, 0_{\mathcal{A}}, 1_{\mathcal{A}})$ consisting of:
    \begin{enumerate}
        \item A set $\mathcal{A}$;
        \item A reflexive and symmetric binary relation $\odot\subseteq\mathcal{A}\times\mathcal{A}$, called the compatibility relation;
        \item A (total) unary operation $\lnot:\mathcal{A}\rightarrow\mathcal{A}$;
        \item A (partial) binary operation $\land:\odot\rightarrow\mathcal{A}$;
        \item The bottom and elements $0_{\mathcal{A}}, 1_{\mathcal{A}} \in\mathcal{A}$,
    \end{enumerate}
     which satisfies that: every subset $S \subseteq\mathcal{A}$ of pairwise compatible elements (i.e., $a \odot b$ for all $a, b \in S$) is contained in some Boolean subalgebra $\mathcal{B} \subseteq\mathcal{A}$, where the operations of $\mathcal{B}$ are defined as the restrictions of $\lnot$ and $\land$ to $\mathcal{B}$.
\end{definition}
\vspace{10pt}

Specially, a Boolean algebra $\mathcal{B}$ is a pBA in which the compatibility relation is defined for all pairs of elements, i.e., $\odot=\mathcal{B}\times\mathcal{B}$.

In the context of physical experiments, the elements of a pBA $\mathcal{A}$ represent observable events. For any $a \in\mathcal{A}$, the negation $\lnot a$ represents the event that $a$ does not occur. For $a, b\in\mathcal{A}$, if $a \odot b$ (i.e., they are compatible or commeasurable), then the conjunction $a \land b$ represents the event that both $a$ and $b$ occur simultaneously. Since $\{\lnot, \land\}$ is an adequate set of connectives, $\mathcal{A}$ encompasses all events that can be observed in an experiment.

For convenience, we define the disjunction operation on a pBA as the De Morgan dual of conjunction:
\[a \lor b := \lnot(\lnot a \land\lnot b)\quad\text{if}\quad a \odot b,\]
which represents the event that either $a$ or $b$ occurs.

Within a pBA $\mathcal{A}$, every Boolean subalgebra naturally represents a \emph{context}: a maximal set of pairwise commeasurable events. This corresponds directly to the notion of a context $C \in\mathcal{M}$ in observable-based theories. Consequently, the pBA $\mathcal{A}$ characterizes the \emph{scenario} of an (generalized) experiment from the event-based perspective.

The structure-preserving maps between partial Boolean algebras are defined as follows.

\vspace{10pt}
\begin{definition}
\label{def-pba_homomorphism}
Let $\mathcal{A}_1$ and $\mathcal{A}_2$ be \rm{pBA}s. A map $f: \mathcal{A}_1 \to\mathcal{A}_2$ is called a \textbf{homomorphism} if it satisfies the following conditions for all $a, b \in\mathcal{A}_1$:
 \begin{enumerate}
    \item $f(0_{\mathcal{A}_1}) = 0_{\mathcal{A}_2}$ and $f(1_{\mathcal{A}_1}) = 1_{\mathcal{A}_2}$.
    \item $f(\neg a) = \neg f(a)$.
    \item If $a \odot b$, then $f(a) \odot f(b)$.
    \item If $a \odot b$, then $f(a \land b) = f(a) \land f(b)$.
 \end{enumerate}
An injective homomorphism is called an \textbf{embedding}. A bijective homomorphism $f$ is called an \textbf{isomorphism} if it also reflects compatibility: $a \odot b$ if and only if $f(a) \odot f(b)$. If such an isomorphism exists, we say $\mathcal{A}_1$ and $\mathcal{A}_2$ are isomorphic, denoted $\mathcal{A}_1 \cong\mathcal{A}_2$.
\end{definition}
\vspace{10pt}

The probability distributions on events in a pBA are described by \emph{states}, which generalize classical probability measures.

\vspace{10pt}
\begin{definition}
\label{def-pba_state}
Let $\mathcal{A}$ be a \rm{pBA}. A \textbf{state} on $\mathcal{A}$ is a function $p: \mathcal{A} \to [0, 1]$ satisfying following conditions for all $a,b\in\mathcal{A}$:
 \begin{enumerate}
    \item $p(0) = 0$ and $p(1) = 1$.
    \item $p(\neg a) = 1 - p(a)$.
    \item $p(a) + p(b) = p(a \land b) + p(a \lor b)$ if $a \odot b$.
 \end{enumerate}
A state $p$ is called \textbf{deterministic} if either $p(a)=0$ or $p(a)=1$ for all $a \in\mathcal{A}$. We denote the set of all states on $\mathcal{A}$ by $s(\mathcal{A})$, and the set of deterministic states by $s_d(\mathcal{A})$.
\end{definition}
\vspace{10pt}

The state on a pBA formally captures the principle of \emph{local consistency} \cite{Abramsky2015Contextuality}, also known as \emph{non-disturbance} \cite{Ramanathan2012Generalized} or \emph{no-signalling} \cite{Popescu1994Quantum}. This means that for any state $p \in s(\mathcal{A})$ and for any two Boolean subalgebras $\mathcal{B}_1, \mathcal{B}_2 \subseteq\mathcal{A}$, the restriction of $p$ to each subalgebra is consistent:  $p|_{\mathcal{B}_1}(e) = p|_{\mathcal{B}_2}(e) = p(e)$ for every event $e \in\mathcal{B}_1 \cap\mathcal{B}_2$.

The deterministic states in $s_d(\mathcal{A})$ are precisely the homomorphisms from the pBA $\mathcal{A}$ to the two-element Boolean algebra $\mathbb{B}_2 = \{0, 1\}$. In logical terms, they correspond to the global truth-value assignments across all contexts.

Next, we introduce the \emph{logical exclusivity principle} \cite{Abramsky2020The} for partial Boolean algebras. This principle will provide several powerful properties to pBA. Firstly, we define two binary relations: the partial order $\leq$ and the exclusivity relation $\perp$. Let $\mathcal{A}$ be a pBA. For $a,b\in \mathcal{A}$, define:
\begin{equation}
\begin{split}
a\leq b&\quad\text{if}\quad a\land b=a.\\
a\perp b&\quad\text{if}\quad a\leq c\text{ and }b\leq\neg c\text{ for some }c\in\mathcal{A}.
\end{split}
\end{equation}

Intuitively, $a \leq b$ means that whenever event $a$ occurs, event $b$ must also occur (i.e., $a$ logically implies $b$). The relation $a \perp b$ signifies that events $a$ and $b$ are mutually exclusive, i.e., they cannot occur simultaneously.

\vspace{10pt}
\begin{definition}
\label{def-LEP}
A partial Boolean algebra $\mathcal{A}$ is said to satisfy the \textbf{logical exclusivity principle (\rm{LEP})} if each pair of exclusive events is compatible, i.e., ${\perp} \subseteq {\odot}$. A \rm{pBA} satisfying \rm{LEP} is called an \textbf{exclusive partial Boolean algebra}, abbreviated by \rm{epBA}.
\end{definition}
\vspace{10pt}

The logical exclusivity principle (LEP) was introduced by Abramsky and Barbosa \cite{Abramsky2020The} to generalize the exclusivity principle from observable-based theories \cite{Adan2012Specker, Fritz2013Local} into the logic framework. This principle serves to filter out pathological pBAs that exhibit undesirable logical behavior.

\begin{figure}[H]
      \begin{minipage}[t]{0.5\linewidth}
          \centering
          \includegraphics[width=0.9\linewidth]{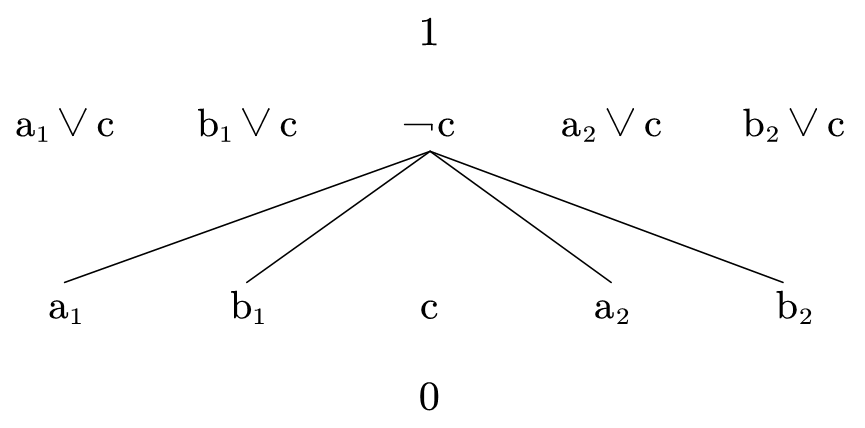}
      \end{minipage}
      \begin{minipage}[t]{0.5\linewidth}
          \includegraphics[width=0.9\linewidth]{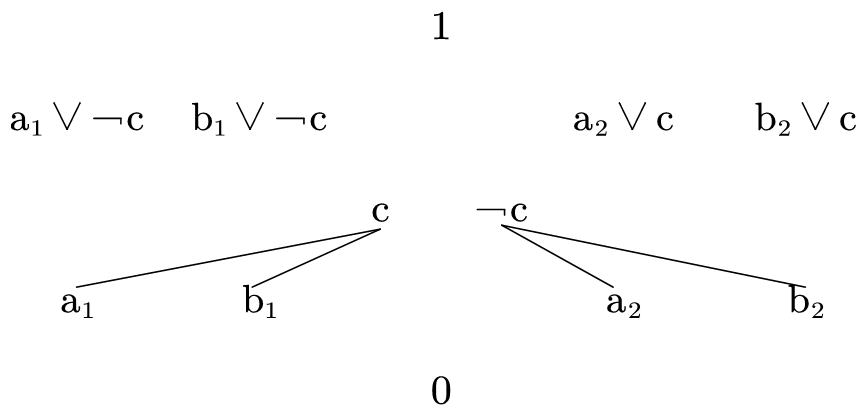}
      \end{minipage}
      \caption{Two illustrative examples of pBA. The left one ($\mathcal{A}_1$) satisfies the LEP and transitivity, while the right one ($\mathcal{A}_2$) violates both.}
    \label{fig:pBAexample}
\end{figure}

For example, consider the two pBAs $\mathcal{A}_1$ and $\mathcal{A}_2$ depicted in Fig.~\ref{fig:pBAexample}, which schematically represent the elements and their partial order relations (note: not all relations are shown). $\mathcal{A}_1$ and $\mathcal{A}_2$ both consist of two 8-element maximal Boolean subalgebras. Those of $\mathcal{A}_1$ are generated by $\{a_1, b_1, c\}$ and $\{a_2, b_2, c\}$ respectively, whereas those of $\mathcal{A}_2$ are generated by $\{a_1, b_1, \neg c\}$ and $\{a_2, b_2, c\}$.

In $\mathcal{A}_2$, we have the relations $a_1 \leq c$ and $c \leq (a_2 \lor c)$. However, we do not have $a_1 \leq (a_2 \lor c)$, because $a_1$ and $a_2 \lor c$ do not reside in a common Boolean subalgebra, in other words, they are incompatible. Consequently, the transitivity of $\leq$ fails in $\mathcal{A}_2$. In contrast, $\mathcal{A}_1$ does not exhibit this pathological behavior.

It is proven in \cite{Abramsky2020The} that a pBA $\mathcal{A}$ satisfies the transitivity if and only if it satisfies LEP. Therefore, LEP is a necessary condition for a pBA to support normal, consistent logical reasoning. It is straightforward to verify that in the example above, $\mathcal{A}_1$ satisfies LEP (and hence transitivity), while $\mathcal{A}_2$ does not.

Henceforth, we focus on the finite exclusive partial Boolean algebra $\mathcal{A}$, which characterizes generalized experimental scenarios satisfying both the no-signalling principle and the logical exclusivity principle. Then the tuple $(\mathcal{A},p)$ characterizes the (static) generalized experiment.

\vspace{10pt}
\begin{definition}
A \textbf{(finite) general system} is a tuple $(\mathcal{A}, p)$, where $\mathcal{A}$ is an (finite) epBA and $p\in s(\mathcal{A})$.
\end{definition}
\vspace{10pt}

A key revelation is that every finite general system is uniquely determined by atoms \cite{Liu2025Atom}. To elaborate on this, we first present the definition of the atom graph.

\vspace{10pt}
\begin{definition}
Let $\mathcal{A}$ be a finite \rm{epBA}.
\begin{itemize}
    \item An \textbf{atom} of $\mathcal{A}$ is a nonzero element $a \in\mathcal{A}$ such that for any $x \in\mathcal{A}$, $x \leq a$ implies $x = 0_{\mathcal{A}}$ or $x = a$. Denote by $\mathrm{At}(\mathcal{A})$ the set of all atoms of $\mathcal{A}$.
    \item The \textbf{atom graph} of $\mathcal{A}$, denoted $\mathcal{G}_a(\mathcal{A})$, is the simple graph with vertex set $\mathrm{At}(\mathcal{A})$, where two distinct vertices $v_1, v_2 \in\mathrm{At}(\mathcal{A})$ are adjacent if and only if $v_1 \odot v_2$.
\end{itemize}
\end{definition}
\vspace{10pt}

In other words, the atom graph has the atoms as its vertices, with edges representing the compatibility relation. For a finite Boolean algebra $\mathcal{B}$, the atom graph $\mathcal{G}_a(\mathcal{B})$ is a complete graph (a clique). Generally, atom graphs of finite epBAs arise from overlapping or gluing together such cliques.

\begin{figure}[h]
    \centering
    \includegraphics[width=0.3\linewidth]{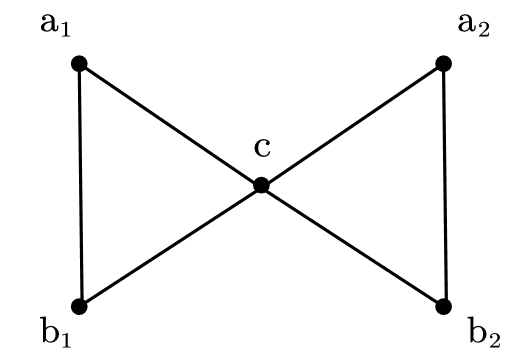}
    \caption{The atom graph $G_a(\mathcal{A}_1)$ of $\mathcal{A}_1$ in Fig~\ref{fig:pBAexample} }\label{GaA1}
 \end{figure}

For example, the atom graph of $\mathcal{A}_1$ in Fig~\ref{fig:pBAexample} is shown in Fig~\ref{GaA1}.

The following results demonstrate that the atom graph completely characterizes the structure and states of epBA.

\vspace{10pt}
\begin{definition}
\label{def-graph_state}
Let $\mathcal{G} = (V, E)$ be a finite simple graph. A \textbf{state} on $\mathcal{G}$ is a function $p: V \to [0, 1]$ such that for every maximal clique $C \subseteq V$, $\sum_{v \in C} p(v) = 1$. The set of all states on $\mathcal{G}$ is denoted by $s(\mathcal{G})$.
\end{definition}
\vspace{10pt}

\begin{theorem}[{\cite{Liu2025Atom}}]
\label{thm:atom_graph_correspondence}
Let $\mathcal{A}$ and $\mathcal{A}'$ be finite \rm{epBA}s. Then:
\begin{enumerate}
    \item There exists a canonical bijection between the states on $\mathcal{A}$ and the states on its atom graph $\mathcal{G}_a(\mathcal{A})$:
          \[\begin{array}{rl}
              \ell: s(\mathcal{A}) &\xrightarrow{\sim} s(\mathcal{G}_a(\mathcal{A})) \\
              p &\mapsto\left. p \right|_{\mathrm{At}(\mathcal{A})}
          \end{array}\]
    \item The \rm{epBA}s are isomorphic if and only if their atom graphs are isomorphic:
          \[\mathcal{A} \cong\mathcal{A}' \quad\text{if and only if}\quad\mathcal{G}_a(\mathcal{A}) \cong\mathcal{G}_a(\mathcal{A}'). \]
          \end{enumerate}
\end{theorem}
\vspace{10pt}

Theorem~\ref{thm:atom_graph_correspondence} establishes that both the algebraic structure and the states of any finite epBA are completely determined by its atom graph. Consequently, experiments described by finite general systems can be equivalently characterized by graphs.

\section{Classical system and quantum system}\label{sec-CS and QS}

This section introduces the use of partial Boolean algebra (pBA) concepts to characterize (static) classical and quantum experiments.

Classical experiments are described within the framework of classical probability theory. Let $\Lambda$ denote the hidden-variable space (i.e., the sample space), $\mathcal{B}$ be a Boolean algebra (more precisely, a $\sigma$-algebra) consisting of subsets of $\Lambda$, and $p$ be a classical probability measure. The triple $(\Lambda, \mathcal{B}, p)$ then forms the axiomatic basis of classical probability theory. Note that each hidden variable $\lambda\in\Lambda$ corresponds a deterministic state on $\mathcal{B}$. Consequently, $\Lambda$ can be depicted by $s_d(\mathcal{B})$, and then the triple $(\Lambda, \mathcal{B}, p)$ can be reduced to the tuple $(\mathcal{B}, p)$, which facilitates a more direct comparison with quantum systems, as quantum mechanics does not admit a hidden-variable space.

\vspace{10pt}
\begin{definition}
A \textbf{(finite) classical system} is a tuple $(\mathcal{B}, p_{\mathcal{B}})$, where $\mathcal{B}$ is a (finite) Boolean algebra and $p_{\mathcal{B}} \in s(\mathcal{B})$.
\end{definition}
\vspace{10pt}

We also call the Boolean algebra $\mathcal{B}$ a \emph{classical scenario}, and $p_{\mathcal{B}}$ a \emph{classical state}.

From the physical perspective, an event corresponds to a proposition of measurement outcomes, such as ``$A \in\Delta$", where $A$ denotes an observable and $\Delta\subseteq\mathbb{R}$ is a Borel set. In quantum mechanics, the observable $A$ is represented by a bounded self-adjoint operator $\hat{A}$, and the event is associated with the spectral projector $\hat{P}_\Delta$ onto the subspace corresponding to eigenvalues in $\Delta$.

Following the construction of standard quantum logic \cite{Birkhoff1936The} and partial Boolean algebra \cite{Kochen1967The}, let $\mathbf{P}(\mathcal{H})$ denote the set of all projectors onto the Hilbert space $\mathcal{H}$. For projectors $\hat{P}, \hat{Q} \in\mathbf{P}(\mathcal{H})$, define the compatibility relation $\odot$ by:
\[\hat{P} \odot\hat{Q} \iff [\hat{P}, \hat{Q}] = \mathbf{0}\]
where $[ \cdot , \cdot ]$ denotes the commutator. When $\hat{P} \odot\hat{Q}$, we say the projectors are compatible.

For any projector $\hat{P}$ onto subspace $S \subseteq\mathcal{H}$, define its orthogonal complement as:
\begin{equation}
\neg\hat{P} := \mathbf{I} - \hat{P}\text{ projecting onto }S^\perp
\end{equation}

 Given compatible projectors $\hat{P} \odot\hat{Q}$ onto subspaces $S_P, S_Q$ respectively, define:
\begin{equation}
\begin{split}
\hat{P} \land\hat{Q} &:= \text{projector onto } S_P \cap S_Q \\
\hat{P} \leq \hat {Q} & \text{ if } \hat{P}\land\hat{Q}=\hat{P}
\end{split}
\end{equation}
with $\mathbf{I}$ the identity operator (projector onto $\mathcal{H}$) and $\mathbf{0}$ the null projector. One can verify that the structure $(\mathbf{P}(\mathcal{H}), \odot, \land, \neg, \mathbf{0}, \mathbf{I})$ constitutes a partial Boolean algebra.

To connect with the standard projector operations, for compatible projectors $\hat{P},\hat{Q} \in\mathbf{P}(\mathcal{H})$, we further have:
\begin{equation}
\begin{split}
\hat{P}\land\hat{Q} &= \hat{P}\hat{Q};\\
\hat{P}\lor\hat{Q} &= \neg(\neg\hat{P}\land\neg\hat{Q}) \\
                    &= \hat{P}+\hat{Q}-\hat{P}\hat{Q}.
\end{split}
\end{equation}

In any practical quantum experiment, only a finite number of events can be observed. These events form a finite partial Boolean subalgebra of $\mathbf{P}(\mathcal{H})$, which introduces the concept of \emph{quantum scenarios}.

\vspace{10pt}
\begin{definition}\label{def-quantum scenario}
A \textbf{(finite) quantum scenario} $\mathcal{Q}$ on $\mathcal{H}$ is a (finite) partial Boolean subalgebra of $\mathbf{P}(\mathcal{H})$.
\end{definition}
\vspace{10pt}

Quantum scenarios model the collection of all observable quantum events in an experiment. For any compatible projectors $\hat{P}, \hat{Q} \in\mathbf{P}(\mathcal{H})$, the events $\neg\hat{P}$, $\neg\hat{Q}$, and $\hat{P} \land\hat{Q}$ can still be observed experimentally. Conversely, if $\hat{P}$ and $\hat{Q}$ are incompatible, the conjunction $\hat{P} \land\hat{Q}$ lacks a clear operational meaning or consistent physical interpretation \cite{Foulis2006Quantum, Kochen2015Reconstruction}.

Furthermore, we have:

\vspace{10pt}
\begin{theorem}
\label{thm-exclusive}
Quantum scenarios are exclusive.
\end{theorem}
\begin{proof}
Let $\mathcal{Q}$ be a quantum scenario. For $\hat{P}_1, \hat{P}_2 \in\mathcal{Q}$, if $\hat{P}_1\perp\hat{P}_2$, i.e., there exists a projector $\hat{Q}\in\mathcal{Q}$ such that $\hat{P}_1 \leq\hat{Q}$ and $\hat{P}_2 \leq\neg\hat{Q}$, then $\hat{P}_1$ and $\hat{P}_2$ are orthogonal, so they are compatible.
\end{proof}
\vspace{10pt}

Given a quantum scenario $\mathcal{Q}$ on a Hilbert space $\mathcal{H}$, a quantum state $\rho$ induces a probability distribution over $\mathcal{Q}$. This is formalized as follows:

\vspace{10pt}
\begin{definition}
Let $\mathcal{Q}$ be a quantum scenario on $\mathcal{H}$. A density operator $\rho$ defines a \textbf{quantum state} on $\mathcal{Q}$ via the relation:
\[\rho(\hat{P}) = \mathrm{tr}(\rho\hat{P}) \quad\text{for } \hat{P}\in\mathcal{Q}.\]
The set of all quantum states on $\mathcal{Q}$ is denoted by $s_q(\mathcal{Q})$.
\end{definition}
\vspace{10pt}

For notational convenience, we use $\rho$ to refer both to the density operator on $\mathcal{H}$ and to the induced quantum state on $\mathcal{Q}$. This unified usage introduces no ambiguity within our framework.

\begin{theorem}
\label{thm-sq subset s}
Quantum states are states. That is, for any quantum scenario $\mathcal{Q}$, we have $s_q(\mathcal{Q})\subseteq s(\mathcal{Q})$.
\end{theorem}
\begin{proof}
Let $\rho\in s_q(\mathcal{Q})$. Then for any $\hat{P} \in\mathcal{Q}$:
\begin{align*}
  \rho(\mathbf{0})&=\mathrm{tr}(\rho\mathbf{0})=\mathrm{tr}(\mathbf{0})=0;  \\
  \rho(\mathbf{I})&=\mathrm{tr}(\rho\mathbf{I})=\mathrm{tr}(\rho)=1;\\
  \rho(\neg\hat{P})&=\mathrm{tr}(\rho(\mathbf{I}-\hat{P}))=\mathrm{tr}(\rho)-\mathrm{tr}(\rho\hat{P})=1-\rho(\hat{P}).
\end{align*}
Moreover, for any compatible pair $\hat{P}, \hat{Q} \in\mathcal{Q}$:
\begin{align*}
\rho(\hat{P}\land\hat{Q})+\rho(\hat{P}\lor\hat{Q})&=
\mathrm{tr}(\rho\hat{P}\hat{Q})+\mathrm{tr}(\rho(\hat{P} + \hat{Q} - \hat{P}\hat{Q}))\\&=
\mathrm{tr}(\rho\hat{P})+\mathrm{tr}(\rho\hat{Q})\\&=
\rho(\hat{P})+\rho(\hat{Q}).
\end{align*}
Hence, $\rho$ satisfies all the conditions in Definition~\ref{def-pba_state}. Thus $\rho\in s(\mathcal{Q})$.
\end{proof}

In summary, the pair $(\mathcal{Q}, \rho)$ fully captures the observable events and their probability distribution in a quantum experiment. This leads naturally to the following definition:

\vspace{10pt}
\begin{definition}
A \textbf{(finite) quantum system} is a tuple $(\mathcal{Q},\rho)$, where $\mathcal{Q}$ is a (finite) quantum scenario and $\rho\in s_q(\mathcal{Q})$.
\end{definition}
\vspace{10pt}

By Theorem~\ref{thm-exclusive} and Theorem~\ref{thm-sq subset s}, quantum systems form a subset of general systems.

Theorem~\ref{thm-sq subset s} shows that for general quantum scenarios $\mathcal{Q}$, we have $s_q(\mathcal{Q})\subseteq s(\mathcal{Q})$. A notable special case arises when $\mathcal{Q} = \mathbf{P}(\mathcal{H})$, the quantum scenario of all projectors on $\mathcal{H}$. In this setting, Gleason's theorem provides a profound characterization of the state space.

\vspace{10pt}
\begin{theorem}[\cite{Gleason1957Measures}]
Suppose $\dim(\mathcal{H})\geq 3$. Then every state on $\mathbf{P}(\mathcal{H})$ arises from a density operator, i.e., $s_q(\mathbf{P}(\mathcal{H})) = s(\mathbf{P}(\mathcal{H}))$.
\end{theorem}
\vspace{10pt}

It is straightforward to demonstrate that Gleason's theorem fails to hold in the 2-dimensional case. Suppose $\dim(\mathcal{H}) = 2$. Then every nontrivial projector $\hat{P} \in\mathbf{P}(\mathcal{H})$ generates exactly one maximal Boolean subalgebra of the form $\{\mathbf{0}, \hat{P}, \neg\hat{P}, \mathbf{I}\}$. These maximal Boolean subalgebras are pairwise disjoint except for the common elements $\mathbf{0}$ and $\mathbf{I}$.

Hence, one can define a state $p \in s(\mathbf{P}(\mathcal{H}))$ such that on each maximal Boolean subalgebra,
\[p(\hat{P}) = 1 \quad\text{and} \quad p(\neg\hat{P}) = 0.\]
However, such a function $p$ is clearly not a quantum state.

\section{Classical embedding, nonclassicality and contextuality}\label{sec-classical embedding}

This section develops a unified framework for analyzing contextuality and nonclassicality of quantum systems.

To determine whether a quantum system $(\mathcal{Q},\rho)$ is classically realizable, one must first identify a classical scenario $\mathcal{B}$ that corresponds to $\mathcal{Q}$, capturing all events and their logical relations. Then two cases arise:
\begin{enumerate}
    \item If no such $\mathcal{B}$ exists, then the quantum scenario $\mathcal{Q}$ itself is nonclassical, such as the Kochen-Specker scenarios \cite{Kochen1967The, Cabello1997Bell, Lisonek2014Kochen}, which will be discussed in Subsection~\ref{subsec-KS}.
    \item If such a $\mathcal{B}$ exists, the next step is to determine whether the quantum state $\rho$ can be represented by a classical state $p$ on $\mathcal{B}$, that is, whether $\rho$ is noncontextual.
\end{enumerate}

This discussion also holds for general systems, formalized in the following definition:

\vspace{10pt}
\begin{definition}\label{def-nc}
A general system $(\mathcal{A}, p)$ is \textbf{classical} if there exists a classical system $(\mathcal{B}, p_{\mathcal{B}})$ and an embedding $i:\mathcal{A}\to\mathcal{B}$ such that
\[p_{\mathcal{B}}(i(e)) = p(e) \quad\text{for all } e\in\mathcal{A}.\]
In this case, we say $p$ is \textbf{noncontextual}.
\end{definition}
\vspace{10pt}

Conversely, $(\mathcal{A}, p)$ is nonclassical if either $\mathcal{A}$ admits no Boolean embedding or $p$ is contextual. Compared with observable-based theories, Definition~\ref{def-nc} accounts for both structures of scenarios and properties of states, thereby clarifying the relationship between nonclassicality and contextuality.

Before proceeding to a deeper analysis of nonclassicality, we demonstrate the equivalence between the definition of noncontextual states in Definition~\ref{def-nc} and the traditional notion of noncontextual hidden-variable models (NCHV) in observable-based theories. We illustrate this connection using the sheaf-theoretic framework (see Appendix~\ref{appendix-A} for details).

Consider a general system $(\mathcal{A}, p)$. Each Boolean subalgebra $\mathcal{B} \subseteq\mathcal{A}$ corresponds to an observable $A_{\mathcal{B}}$, where the elements of $\mathcal{B}$ represent the possible outcomes of measuring $A_{\mathcal{B}}$. Two observables $A_{\mathcal{B}_1}$ and $A_{\mathcal{B}_2}$ are compatible if and only if $\mathcal{B}_1$ and $\mathcal{B}_2$ are both contained in a common Boolean subalgebra $\mathcal{B}_3$. In this case, $\mathcal{B}_3$ represents the context $\{A_{\mathcal{B}_1}, A_{\mathcal{B}_2}\}$ and likewise corresponds to their joint measurement $A_{\mathcal{B}_1}A_{\mathcal{B}_2}$. Hence, the family of Boolean subalgebras of $\mathcal{A}$ characterizes the observables and contexts. This leads to the following result:

\vspace{10pt}
\begin{theorem}\label{thm-contextuality}
Let $(\mathcal{A}, p)$ be a general system. The state $p$ is noncontextual if and only if it admits a NCHV model.
\end{theorem}
\begin{proof}
By Theorem~\ref{thm-sheaf-nc}, it suffices to show that $p$ corresponds to a noncontextual no-signaling empirical model $\{p_C\}$.

Let $\{\mathcal{B}_i\}_{i \in I}$ be the family of all Boolean subalgebras of $\mathcal{A}$, where $I$ is an index set. Each $\mathcal{B}_i$ corresponds to a context $C_i$, and its elements represent joint outcomes on $C_i$. The state $p$ thus induces an empirical model $\{p_{C_i} : i \in I\}$ defined by $p_{C_i}(e) = p(e)$ for all $e \in\mathcal{B}_i$.

Suppose $C_i \subseteq C_j$, which implies $\mathcal{B}_i \subseteq\mathcal{B}_j$. Then for any $e \in\mathcal{B}_i$, we have $p_{C_i}(e) = p(e) = p_{C_j}(e)$, so $p_{C_i} = p_{C_j}|_{C_i}$. Hence, $\{p_{C_i} : i \in I\}$ is no-signaling.

Since $p$ is noncontextual, $\mathcal{A}$ can be embedded into a Boolean algebra $\mathcal{B}$. For convenience, we treat $\mathcal{A}$ as a partial Boolean subalgebra of $\mathcal{B}$, and each $\mathcal{B}_i$ as a Boolean subalgebra of $\mathcal{B}$. Then $\mathcal{B}$ corresponds to a global context $C$ containing every $C_i$ ($i \in I$). Moreover, there exists a classical state $p_C$ on $\mathcal{B}$ such that $p_C(e) = p(e)$ for all $e \in\mathcal{A}$, which implies $p_C|_{C_i} = p_{C_i}$ for all $i \in I$. Therefore, $\{p_{C_i} : i \in I\}$ is noncontextual.
\end{proof}
\vspace{10pt}

To determine whether a general system $(\mathcal{A}, p)$ is classical, one must verify whether $\mathcal{A}$ can be embedded into some Boolean algebra $\mathcal{B}$. Intuitively, a larger $\mathcal{B}$ may facilitate such an embedding, but it also complicates theoretical analysis. Nevertheless, we can establish a unified and minimal corresponding Boolean algebra for each $\mathcal{A}$.

From the perspective of hidden-variable theory, $(\mathcal{A}, p)$ is classical if there exists a hidden-variable space $\Lambda$ such that the occurrence of every event $e \in\mathcal{A}$ is determined by some $\lambda\in\Lambda$. Mathematically, this means $\mathcal{A}$ can be embedded into the power set $\mathcal{P}(\Lambda)$ via the map:
\begin{equation}
\begin{split}
i:\ &\mathcal{A}\to\mathcal{P}(\Lambda)\\
&e\mapsto\{\lambda|\ \lambda(e)=1\}
\end{split}
\end{equation}
Conversely, each hidden variable $\lambda\in\Lambda$ induces a deterministic state on $\mathcal{A}$. For finite general systems, we prove that if the hidden-variable space exists, it is captured precisely by the set $s_d(\mathcal{A})$ of deterministic states.

\vspace{10pt}
\begin{lemma}\label{lemma1}
Let $\mathcal{A}$ be a finite epBA and $a, b \in\mathcal{A}$. If $\mathcal{A}$ can be embedded into a Boolean algebra $\mathcal{B}$, and $\lambda(a) = \lambda(b)$ for all $\lambda\in s_d(\mathcal{A})$, then $a = b$.
\end{lemma}
\begin{proof}
Since $\mathcal{A}$ is finite, it generates a finite Boolean subalgebra of $\mathcal{B}$. Without loss of generality, assume $\mathcal{B}$ is finite. Then we may take $\mathcal{B} = \mathcal{P}(S)$ for some finite set $S$.

Let $i : \mathcal{A} \to\mathcal{P}(S)$ be an embedding. For each $s \in S$, define the state $\lambda_s(e) = 1$ if $s \in i(e)$, and $\lambda_s(e) = 0$ otherwise. It is straightforward to verify that $\lambda_s \in s_d(\mathcal{A})$. By assumption, $\lambda_s(a) = \lambda_s(b)$ for all $s \in S$, which implies $s \in i(a)$ if and only if $s \in i(b)$. Hence $i(a) = i(b)$, and since $i$ is injective, $a = b$.
\end{proof}
\vspace{10pt}

\begin{theorem}\label{thm-imbed}
Let $\mathcal{A}$ be a finite epBA. Then $\mathcal{A}$ can be embedded into a Boolean algebra if and only if it can be embedded into $\mathcal{P}(s_d(\mathcal{A}))$.
\end{theorem}
\begin{proof}
Sufficiency is clear. For necessity, assume $\mathcal{A}$ embeds into some Boolean algebra. Then $s_d(\mathcal{A})$ is nonempty. Define the map:
\begin{equation}
\label{eq-classical embedding}
\begin{split}
i_{\mathcal{A}}:\ &\mathcal{A}\to\mathcal{P}(s_d(\mathcal{A}))\\
&e\mapsto\{\lambda\in s_d(\mathcal{A}): \lambda(e)=1\}
\end{split}
\end{equation}
By Lemma~\ref{lemma1}, if $a \neq b$, there exists $\lambda\in s_d(\mathcal{A})$ such that $\lambda(a) \neq\lambda(b)$, so $i_{\mathcal{A}}(a) \neq i_{\mathcal{A}}(b)$. Hence $i_{\mathcal{A}}$ is injective.

We now show that $i_{\mathcal{A}}$ is a homomorphism:
\begin{enumerate}
    \item$i_{\mathcal{A}}(0) = \{\lambda\in s_d(\mathcal{A}):\lambda(0)=1\}=\emptyset$ and $i_{\mathcal{A}}(1) = \{\lambda\in s_d(\mathcal{A}):\lambda(1)=1\}=s_d(\mathcal{A})$;
    \item$i_{\mathcal{A}}(\neg e) = \{\lambda\in s_d(\mathcal{A}):\lambda(\neg e)=1\}=\{\lambda\in s_d(\mathcal{A}):\lambda(e)=0\}=s_d(\mathcal{A})\setminus i_{\mathcal{A}}(e) = \neg i_{\mathcal{A}}(e)$;
    \item For compatible $a, b \in\mathcal{A}$, $i_{\mathcal{A}}(a)$ and $i_{\mathcal{A}}(b)$ are compatible;
    \item For compatible $a, b \in\mathcal{A}$: $i_{\mathcal{A}}(a \land b) = \{\lambda\in s_d(\mathcal{A}):\lambda(a\land b)=1\}=\{\lambda\in s_d(\mathcal{A}): \lambda(a)=1\}\cap\{\lambda\in s_d(\mathcal{A}): \lambda(b)=1\} =i_{\mathcal{A}}(a) \cap i_{\mathcal{A}}(b)$.
\end{enumerate}
Therefore, $i_{\mathcal{A}}$ is an embedding from $\mathcal{A}$ into $\mathcal{P}(s_d(\mathcal{A}))$.
\end{proof}
\vspace{10pt}

We refer to the embedding $i_{\mathcal{A}}$ defined in \eqref{eq-classical embedding} as the \emph{classical embedding} of $\mathcal{A}$. To simplify notation, we introduce the following conventions:
\begin{equation}
\begin{split}
\mathcal{A}^c &:= \mathcal{P}(s_d(\mathcal{A})); \\
e^c &:= i_{\mathcal{A}}(e) \quad\text{for all } e \in\mathcal{A}.
\end{split}
\end{equation}

Next we prove that $\mathcal{A}^c$ is the minimal Boolean algebra that admits an embedding from $\mathcal{A}$.

\vspace{10pt}
\begin{lemma}\label{p-monotonicity}
Let $\mathcal{A}$ be a pBA and $p \in s(\mathcal{A})$ a state. For any $a, b \in\mathcal{A}$, if $a \leq b$, then $p(a) \leq p(b)$.
\end{lemma}
\begin{proof}
Assume $a \leq b$, which means $a = a \land b$. Then:
\[
a \lor b = (a \land b) \lor b = (a \lor b) \land (b \lor b) = (a \lor b) \land b,
\]
so $a \lor b \leq b$, and thus $a \lor b = b$. Therefore,
\[
p(b) = p(a \lor b) = p(a \lor (b \land\neg a)) = p(a) + p(b \land\neg a) - p(a \land b \land\neg a) = p(a) + p(b \land\neg a) \geq p(a),
\]
which completes the proof.
\end{proof}
\vspace{10pt}

\begin{lemma}\label{lemma2}
Let $\mathcal{A}$ be a finite epBA, and let $i: \mathcal{A} \to\mathcal{B}$ be an embedding into a Boolean algebra $\mathcal{B}$. For any two distinct deterministic states $\lambda_1, \lambda_2 \in s_d(\mathcal{A})$, the following hold:
\begin{enumerate}
    \item$\bigwedge\limits_{\lambda_1(e) = 1} i(e) \neq\bigwedge\limits_{\lambda_2(e) = 1} i(e)$;
    \item$\bigwedge\limits_{\lambda_1(e) = 1} i(e) \land\bigwedge\limits_{\lambda_2(e) = 1} i(e) = 0_{\mathcal{B}}$.
\end{enumerate}
\end{lemma}
\begin{proof}
Recalling the Definition~\ref{def-pba_state}, easily prove that $\lambda\in s_d(\mathcal{A})$ if and only if for every Boolean subalgebra $\mathcal{C} \subseteq\mathcal{A}$, the restriction $\lambda|_{\mathcal{C}}$ is a deterministic state on $\mathcal{C}$.

We first show that for any $\lambda\in s_d(\mathcal{A})$, if a map $d: \mathcal{A} \to\{0,1\}$ satisfies
\[\{ e \in\mathcal{A} : d(e) = 1 \}\supsetneq\{ e \in\mathcal{A} : \lambda(e) = 1 \},\]
then $d \notin s_d(\mathcal{A})$. Suppose, for contradiction, that such a $d$ were a deterministic state. Then there exists a Boolean subalgebra $\mathcal{C} \subseteq\mathcal{A}$ such that
\[\{ e \in\mathcal{C} : d|_{\mathcal{C}}(e) = 1 \}\supsetneq\{ e \in\mathcal{C} : \lambda|_{\mathcal{C}}(e) = 1 \}.\]
However, each deterministic state on $\mathcal{C}$ corresponds to an atom. Let $\lambda|_{\mathcal{C}}$ correspond to $a_{\lambda}\in\mathcal{C}$. Since $d|_{\mathcal{C}}(a_\lambda) = 1$, if $d|_{\mathcal{C}}\in s_d(\mathcal{C})$, then we have $d|_{\mathcal{C}} = \lambda|_{\mathcal{C}}$, a contradiction. Hence, $d|_{\mathcal{C}}$ is not a deterministic state, and thus $d \notin s_d(\mathcal{A})$.

Now, since $\lambda_1 \neq\lambda_2$, we have
\[\{ e \in\mathcal{A} : \lambda_1(e) = 1 \}\neq\{ e \in\mathcal{A} : \lambda_2(e) = 1 \}.
\]
Suppose, for contradiction, that
\[\bigwedge_{\lambda_1(e) = 1} i(e) = \bigwedge_{\lambda_2(e) = 1} i(e) = b.
\]
Then there exists a deterministic state $\lambda'_b \in s_d(\mathcal{B})$ such that $\lambda'_b(b) = 1$. This state induces a deterministic state $\lambda_b \in s_d(\mathcal{A})$ defined by $\lambda_b(e) = \lambda'_b(i(e))$ for all $e \in\mathcal{A}$, thus satisfying $\lambda_b(e) = 1$ for all $e$ such that $\lambda_1(e) = 1$ or $\lambda_2(e) = 1$. However, this contradicts to $\lambda_b\in s_d(\mathcal{A})$ derived from the earlier result.

For the second claim, consider any $v \in s_d(\mathcal{B})$. This induces a state $\lambda_v \in s_d(\mathcal{A})$ via $\lambda_v(e) = v(i(e))$. Since $\lambda_1 \neq\lambda_2$, there exists some $e \in\mathcal{A}$ such that, without loss of generality, $\lambda_1(e) = 1$ (or $\lambda_2(e) = 1$) but $\lambda_v(e) = 0$. Then
\[
v\left( \bigwedge_{\lambda_1(e)=1} i(e) \land\bigwedge_{\lambda_2(e)=1} i(e) \right) = \prod_{\lambda_1(e)=1} v(i(e)) \cdot\prod_{\lambda_2(e)=1} v(i(e)) = 0,
\]
since the conjunction contains a term $v(i(e)) = 0$. As this holds for all $v \in s_d(\mathcal{B})$, we conclude
\[\bigwedge_{\lambda_1(e)=1} i(e) \land\bigwedge_{\lambda_2(e)=1} i(e) = 0_{\mathcal{B}},
\]
by Lemma~\ref{lemma1}.
\end{proof}
\vspace{10pt}

\begin{theorem}\label{thm-minimal}
Let $\mathcal{A}$ be a finite epBA. If $\mathcal{A}$ can be embedded into a Boolean algebra $\mathcal{B}$, then there exists an embedding from $\mathcal{A}^c$ into $\mathcal{B}$.
\end{theorem}
\begin{proof}
Let $i: \mathcal{A} \to\mathcal{B}$ be an embedding. Define a map $i': \mathcal{A}^c \to\mathcal{B}$ as follows:
\[\begin{aligned}
&i'(\emptyset)= 0_{\mathcal{B}}; \\
&i'(\{\lambda\})=\bigwedge_{\lambda(e) = 1} i(e); \\
&i'(S)=\bigvee_{\lambda\in S} i'(\{\lambda\}),\quad S\subseteq s_d(\mathcal{A}).
\end{aligned}
\]

We now verify that $i'$ is a Boolean algebra homomorphism and is injective.

\begin{enumerate}
    \item
    For any $v \in s_d(\mathcal{B})$, let $\lambda_v \in s_d(\mathcal{A})$ be the state induced by $v$, i.e., $v(i(e)) = \lambda_v(e)$ for all $e \in\mathcal{A}$. Then:
    \[
    i'(s_d(\mathcal{A})) = \bigvee_{\lambda\in s_d(\mathcal{A})} \bigwedge_{\lambda(e)=1} i(e) \geq\bigwedge_{\lambda_v(e)=1} i(e) = \bigwedge_{v(i(e))=1} i(e).
    \]
    Hence, for every $v \in s_d(\mathcal{B})$:
    \[
    v(i'(s_d(\mathcal{A}))) \geq v\left( \bigwedge_{v(i(e))=1} i(e) \right)= \prod_{v(i(e))=1} v(i(e)) = 1.
    \]
    By Lemma~\ref{lemma1}, it follows that $i'(s_d(\mathcal{A})) = 1_{\mathcal{B}}$.

    \item
    From the definition of $i'$:
    \[
    i'(\neg S) \lor i'(S) = \left( \bigvee_{\lambda\notin S} i'(\{\lambda\}) \right) \lor\left( \bigvee_{\lambda\in S} i'(\{\lambda\}) \right) = i'(s_d(\mathcal{A})) = 1_{\mathcal{B}}.
    \]
    Moreover, by Lemma~\ref{lemma2}, for distinct $\lambda_1, \lambda_2 \in s_d(\mathcal{A})$:
    \[
    i'(\{\lambda_1\}) \land i'(\{\lambda_2\}) = 0_{\mathcal{B}}.
    \]
    Therefore, $i'(\neg S) = \neg i'(S)$.

    \item
    By Lemma~\ref{lemma2}:
    \[
    i'(S \cap T) = \bigvee_{\lambda\in S \cap T} i'(\{\lambda\}) = \left( \bigvee_{\lambda\in S} i'(\{\lambda\}) \right) \land\left( \bigvee_{\lambda\in T} i'(\{\lambda\}) \right) = i'(S) \land i'(T).
    \]\end{enumerate}

Thus, $i'$ is a Boolean algebra homomorphism.

To show injectivity, suppose $S \neq T$. Then $S \cap T \subsetneq S$ and $S \cap T \subsetneq T$, so:
\[
i'(S \cap T) = i'(S) \land i'(T) < i'(S) \quad\text{and} \quad i'(T),
\]
which implies $i'(S) \neq i'(T)$. Hence, $i'$ is injective.
\end{proof}
\vspace{10pt}

If $\mathcal{A}$ is an epBA, Theorems~\ref{thm-imbed} and~\ref{thm-minimal} together imply that $\mathcal{A}^c$ is the minimal classical counterpart of $\mathcal{A}$.

Apart from the scenarios, we also provide a unified description of noncontextual states. Let $\mathcal{A}$ be an epBA. We define the convex hull of $s_d(\mathcal{A})$ as:
\begin{equation}
\begin{split}
s_{nc}(\mathcal{A}) &:= \operatorname{conv}\big(s_d(\mathcal{A})\big) \\
&= \left\{\sum_{\lambda\in s_d(\mathcal{A})} k_{\lambda} \lambda : k_{\lambda} \geq 0,\ \sum_{\lambda\in s_d(\mathcal{A})} k_{\lambda} = 1 \right\}.
\end{split}
\end{equation}

Then we have:

\vspace{10pt}
\begin{theorem}\label{thm-convex}
Let $(\mathcal{A}, p)$ be a finite general system. If $\mathcal{A}$ can be embedded into a Boolean algebra, then $p$ is noncontextual if and only if $p \in s_{nc}(\mathcal{A})$.
\end{theorem}
\begin{proof}
Sufficiency: Suppose $p \in s_{nc}(\mathcal{A})$, so that $p = \sum_{\lambda\in s_d(\mathcal{A})} k_{\lambda} \lambda$ with $k_{\lambda} \geq 0$ and $\sum_{\lambda} k_{\lambda} = 1$. By Theorem~\ref{thm-imbed}, $\mathcal{A}$ admits the classical embedding $i_{\mathcal{A}}$. Define a state $p'$ on $\mathcal{A}^c$ by:
\begin{align*}
p'(\emptyset) &= 0; \\
p'(S) &= \sum_{\lambda\in S} k_{\lambda} \quad\text{for } S \subseteq s_d(\mathcal{A}).
\end{align*}
It is straightforward to verify that $p' \in s(\mathcal{A}^c)$. Moreover, for any $e \in\mathcal{A}$:
\[
p'(i_{\mathcal{A}}(e)) = p'(e^c) = \sum_{\lambda\in e^c} k_{\lambda} = \sum_{\lambda(e)=1} k_{\lambda} = p(e).
\]
Hence, $p$ is noncontextual.

Necessity: Suppose $p$ is noncontextual. Then there exists a classical system $(\mathcal{B}, p_{\mathcal{B}})$ satisfying the conditions in Definition~\ref{def-nc}. By Theorems~\ref{thm-imbed} and~\ref{thm-minimal}, we may assume without loss of generality that the classical system is $(\mathcal{A}^c, p')$, and that $p(e) = p'(e^c)$ for all $e \in\mathcal{A}$. It follows that:
\[
p = \sum_{\lambda\in s_d(\mathcal{A})} p'(\{\lambda\}) \lambda,
\]
and therefore $p \in s_{nc}(\mathcal{A})$.
\end{proof}
\vspace{10pt}

Theorem~\ref{thm-convex} shows that noncontextual states are precisely the convex combinations of deterministic states. A similar argument appears in the graph-theoretic approach to contextuality \cite{Adan2014Graph}.

Based on the above, we obtain a unified characterization of classicality:

\vspace{10pt}
\begin{corollary}\label{col-classicality}
A finite general system $(\mathcal{A}, p)$ is classical if and only if $\mathcal{A}$ can be embedded into $\mathcal{A}^c$ and $p \in s_{nc}(\mathcal{A})$.
\end{corollary}
\vspace{10pt}

In other words, $(\mathcal{A}, p)$ is nonclassical if $\mathcal{A}$ cannot be embedded into $\mathcal{A}^c$ or if $p \notin s_{nc}(\mathcal{A})$. This shows that contextuality is a sufficient condition for nonclassicality. Indeed, if $\mathcal{A}$ does not admit a classical embedding, then $\mathcal{A}$ itself exhibits nonclassicality, even in the absence of contextual states. Furthermore, we will present an example in subsection~\ref{subset-NC without C} demonstrating that contextuality is not necessary for nonclassicality.

\section{Quantum nonclassicality and contextuality}\label{sec-QNC}

Applying Corollary~\ref{col-classicality} to finite quantum systems $(\mathcal{Q},\rho)$ yields a unified framework for characterizing quantum nonclassicality and contextuality, which can be classified into three types:
\begin{enumerate}
    \item The quantum scenario $\mathcal{Q}$ cannot be embedded into any Boolean algebra, such as the \emph{Kochen-Specker scenarios} \cite{Kochen1967The, Cabello1997Bell, Lisonek2014Kochen};
    \item$\mathcal{Q}$ admits a classical embedding, yet all quantum states on $\mathcal{Q}$ are contextual, referred to as \emph{state-independent contextuality} (SIC) \cite{Yu2012State};
    \item$\mathcal{Q}$ admits a classical embedding, and some specific quantum state $\rho$ on $\mathcal{Q}$ is contextual, referred to as \emph{state-dependent contextuality} (SDC) or simply contextuality \cite{Clauser1969Proposed, Alexander2008Simple}.
\end{enumerate}

This section formalizes these major types of quantum nonclassicality and contextuality within the epBA framework, and demonstrates how to witness nonclassicality and contextuality using fewer projectors and observables. To this end, we first introduce the notion of \emph{generators} of scenarios.

\vspace{10pt}
\begin{notation}
Let $\mathcal{A}$ be an epBA. If $\mathcal{A}$ is generated by a subset $S = \{e_1, \dots, e_n\}$ under the operations $\neg$ and $\land$, we write $\mathcal{A} = \langle S \rangle$.
\end{notation}
\vspace{10pt}

In practice, if $\langle S \rangle = \mathcal{A}$, then observing the entire scenario $\mathcal{A}$ reduces to observing $S$. The size of the generating set $S$ is typically much smaller than that of $\mathcal{A}$, providing an economical way to characterize scenarios.

For deterministic states $\lambda\in s_d(\mathcal{A})$, we have $\lambda(\neg e) = 1 - \lambda(e)$ for all $e \in\mathcal{A}$, and $\lambda(a \land b) = \lambda(a)\lambda(b)$ whenever $a \odot b$. In other words, the truth values of the generators $S$ determine the truth values of all elements in $\mathcal{A}$. Therefore, when studying scenarios or noncontextual states, we can reduce the problem to considering the generators. However, for general states $p \in s(\mathcal{A})$, not all generating sets are suitable, since the probabilities $p(a)$ and $p(b)$ generally do not determine $p(a \land b)$. This motivates the following definition:

\vspace{10pt}
\begin{definition}
Let $\mathcal{A}$ be an epBA. A subset $S \subseteq\mathcal{A}$ is called a \textbf{perfect generating set} of $\mathcal{A}$ if $\langle S \rangle = \mathcal{A}$ and every state $p \in s(\mathcal{A})$ is uniquely determined by its restriction $p|_S$. More precisely, the restriction map $*|_S : s(\mathcal{A}) \to\{ p|_S : p \in s(\mathcal{A}) \}$ is injective.
\end{definition}
\vspace{10pt}

If $S$ is a perfect generating set of $\mathcal{A}$, then the general system $(\mathcal{A}, p)$ is fully characterized by $(S, p|_S)$. For example, by Theorem~\ref{thm:atom_graph_correspondence}, the set of atoms $\mathrm{At}(\mathcal{A})$ forms a perfect generating set of $\mathcal{A}$. Clearly, any \emph{minimal} set of events witnessing contextuality must be a perfect generating set.

\subsection{Kochen-Specker scenarios}\label{subsec-KS}

This subsection introduces the \emph{Kochen-Specker scenario}, and we show that 12 projectors suffice to generate a Kochen-Specker scenario.

The nonclassicality of certain quantum experiments arises not from contextual states but from the structure of the scenario itself. A prominent example is provided by Kochen-Specker scenarios, which were originally introduced to describe scenarios that admit no global truth-value assignment \cite{Kochen1967The}.

\vspace{10pt}
\begin{definition}
A quantum scenario $\mathcal{Q}$ is called a \textbf{Kochen-Specker scenario} if $s_d(\mathcal{Q}) = \emptyset$, or equivalently, if there exists no homomorphism from $\mathcal{Q}$ to the two-element Boolean algebra $\mathbb{B}_2$.
\end{definition}
\vspace{10pt}

If $\mathcal{Q}$ is a Kochen-Specker scenario, then every quantum system $(\mathcal{Q}, \rho)$ is nonclassical, because every nontrivial Boolean algebra admits homomorphisms to $\mathbb{B}_2$. Hence, $\mathcal{Q}$ cannot be embedded into any Boolean algebra.

In the works of quantum contextuality, Kochen-Specker scenarios are often represented by \emph{Kochen-Specker sets}, sets of vectors that do not admit any global truth-value assignment.

\vspace{10pt}
\begin{definition}
A set $P = \{v_i\}_{i=1}^n$ of vectors in $\mathbb{C}^d$ is called a \textbf{Kochen-Specker set} if there exists no function $f : P \to\{0,1\}$ (called a \textbf{Kochen-Specker assignment}) such that:
\begin{itemize}
    \item$f(v_i)f(v_j) = 0$ whenever $v_i$ and $v_j$ are orthogonal;
    \item$\sum_{i=1}^d f(v_i) = 1$ for every orthonormal basis $\{v_1, \dots, v_d\}\subseteq P$.
\end{itemize}
\end{definition}
\vspace{10pt}

\begin{figure}[H]
    \centering\includegraphics[width=0.45\linewidth]{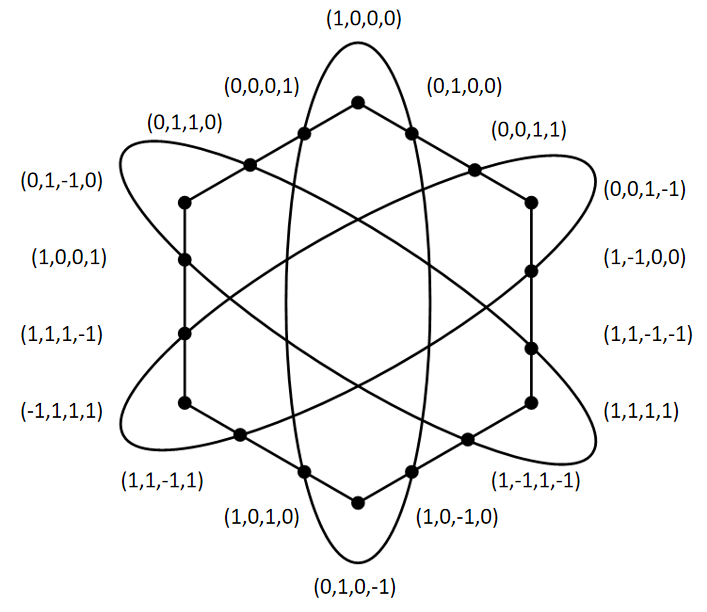}
    \caption{Orthogonality graph of the CEG set (normalization factors omitted). Each straight line or circle represents a maximal clique.}
    \label{CEG}
\end{figure}

Each vector $\ket{\psi}$ corresponds canonically to the rank-1 projector $\ket{\psi}\!\bra{\psi}$, so a Kochen-Specker set can equivalently be viewed as a set of projectors. A well-known example is the Kochen-Specker set constructed by Cabello, Estebaranz, and Garc\'{\i}a-Alcaine (CEG) \cite{Cabello1997Bell}, whose orthogonality graph is shown in Fig.~\ref{CEG}.

The CEG set consists of 18 vectors $\{v_i\}_{i=1}^{18}$ forming 9 orthonormal bases. Suppose, for contradiction, that a Kochen-Specker assignment $f$ exists. Then for each maximal clique $\{v_{k_1}, v_{k_2}, v_{k_3}, v_{k_4}\}$ in Fig.~\ref{CEG}, we must have:
\[
f(v_{k_1}) + f(v_{k_2}) + f(v_{k_3}) + f(v_{k_4}) = 1.
\]
Summing over all 9 cliques yields:
\[
2 \left( \sum_{i=1}^{18} f(v_i) \right) = 9.
\]
The left-hand side is even, while the right-hand side is odd, a contradiction. Therefore, the CEG set is a Kochen-Specker set.

The problem of minimizing the size of Kochen-Specker scenarios has remained a central challenge for decades. The original Kochen-Specker set constructed by Kochen and Specker comprised 117 vectors \cite{Kochen1967The}. The CEG set, with 18 vectors, has been shown to be a minimal Kochen-Specker set \cite{Xu2020Proof}. In three dimensions, such sets are often referred to as \emph{Kochen-Specker systems}. Although the smallest such system remains unknown, it must contain at least 23 vectors \cite{Li2022An}.

\begin{figure}[H]
    \centering\includegraphics[width=0.4\linewidth]{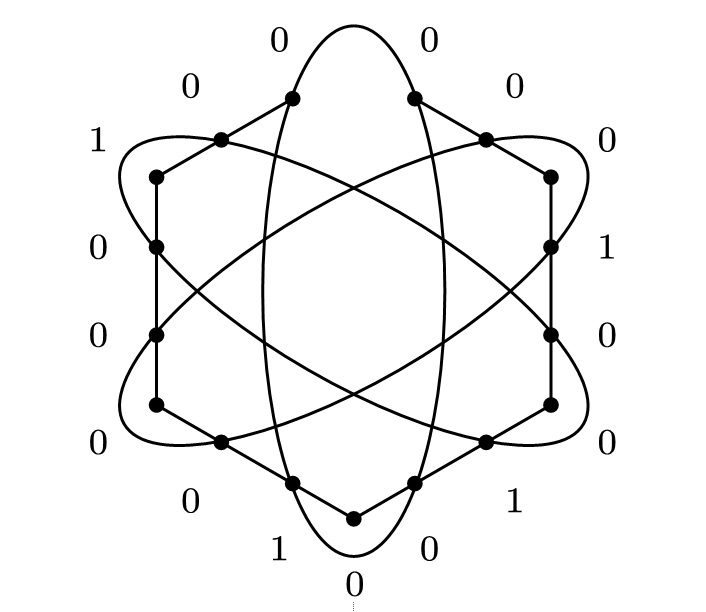}
    \caption{A Kochen--Specker assignment on the 17-vector set CEG$'$.}
    \label{CEG17}
\end{figure}

We emphasize that the logical (event-based) perspective allows the construction of Kochen-Specker scenarios using fewer projectors. For instance, by removing the vector $(1,0,0,0)$ from the CEG set, we obtain a 17-vector set, denoted CEG$'$. Both CEG and CEG$'$ generate the same Kochen-Specker scenario $\mathcal{Q}_{\mathrm{CEG}}$. However, CEG$'$ is not itself a Kochen-Specker set, as it admits a Kochen-Specker assignment, illustrated in Fig.~\ref{CEG17}.

Note that the assignment in Fig.~\ref{CEG17} violates local consistency: it forces the removed vector to be assigned both 0 and 1 across different contexts. This example shows that the nonexistence of a Kochen-Specker assignment is strictly stronger than the nonexistence of a deterministic state. Hence, it is possible to construct a Kochen-Specker scenario using a set of projectors smaller than the CEG set.

\vspace{10pt}
\begin{theorem}\label{thm-12CEG}
The Kochen-Specker scenario $\mathcal{Q}_{\mathrm{CEG}}$ can be generated by $12$ projectors.
\end{theorem}
\begin{proof}
The 18 vectors in the CEG set are contained in 6 maximal cliques. Each clique corresponds to a maximal Boolean subalgebra of $\mathcal{Q}_{\mathrm{CEG}}$ with $2^4$ elements. Since any $2^4$-element Boolean algebra $\mathcal{P}(\{a,b,c,d\})$ is generated by two elements (for instance, $\{a,b\}$ and $\{b,c\}$), the entire algebra $\mathcal{Q}_{\mathrm{CEG}}$ can be generated by 12 projectors: two from each maximal Boolean subalgebra.

As an example, consider the context with atoms
\[\left\{\hat{P}_{(1,0,0,0)},\ \hat{P}_{(0,1,0,0)},\ \hat{P}_{(0,0,1,1)},\ \hat{P}_{(0,0,1,-1)} \right\}.
\]
This Boolean subalgebra can be generated by the two projectors:
\[\hat{P} = \hat{P}_{(1,0,0,0)} \lor\hat{P}_{(0,1,0,0)} \quad\text{and} \quad\hat{Q} = \hat{P}_{(0,1,0,0)} \lor\hat{P}_{(0,0,1,1)},
\]
where $\hat{P}_v$ denotes the projector corresponding to vector $v$.
\end{proof}
\vspace{10pt}

Therefore, the minimal number of projectors required to generate a Kochen-Specker scenario is at most 12.

\subsection{Nonclassicality without contextuality}\label{subset-NC without C}

Based on upon the notions in the previous subsection, we employ a Kochen-Specker set to construct an example of a nonclassical quantum system that admits a noncontextual state.

If a quantum scenario $\mathcal{Q}$ can be embedded into a Boolean algebra, then $s_d(\mathcal{Q})$ is nonempty. However, the converse is not true. Specifically, one can construct a quantum scenario that admits deterministic states while refusing any embedding into a Boolean algebra. A similar idea has been used by \cite{Cabello2015Necessary} in the study of state-independent contextuality.

\vspace{10pt}
\begin{theorem}\label{thm-01vsKS}
There exists a quantum scenario $\mathcal{Q}$ such that $s_d(\mathcal{Q}) \neq\emptyset$, yet $\mathcal{Q}$ cannot be embedded into any Boolean algebra.
\end{theorem}
\begin{proof}
Let $K$ be a Kochen-Specker set in a $d$-dimensional Hilbert space $\mathcal{H}$ (for example, the CEG set in $\mathbb{C}^4$). Denote by $k_i$ the $i$-th component of a vector $\ket{k} \in K$. Define an extended set in $\mathbb{C}^{d+1}$ by:
\[
K' = \left\{ (k_1, \dots, k_d, 0)^T:\ket{k} \in K \right\}\cup\left\{\ket{k'} = (0, \dots, 0, 1)^T\right\}.
\]
We identify $K'$ with the corresponding set of rank-1 projectors. Let $\mathcal{Q} = \langle K' \rangle$ be the quantum scenario generated by $K'$. Then $s_d(\mathcal{Q})$ is nonempty, because we have a deterministic state $\lambda$ defined by
\[\lambda(\ket{k'}\!\bra{k'}) = 1, \qquad\lambda(\hat{P}_{(k_1, \dots, k_d, 0)}) = 0 \quad\text{for all } \ket{k} \in K,
\]
corresponding to the quantum state $\ket{k'}$.

Now suppose, for contradiction, that there exists an embedding $i: \mathcal{Q} \to\mathcal{B}$ into a Boolean algebra. By Lemma~\ref{lemma1}, $i(\ket{k'}\!\bra{k'}) \neq 1_{\mathcal{B}}$. Hence, there exists a deterministic state $v \in s_d(\mathcal{B})$ such that $v(i(\ket{k'}\!\bra{k'})) = 0$. This induces a state $\lambda_v \in s_d(\mathcal{Q})$ with $\lambda_v(\ket{k'}\!\bra{k'}) = 0$. The restriction of $\lambda_v$ to the projectors
\[\left\{\hat{P}_{(k_1, \dots, k_d, 0)}:\ket{k} \in K \right\}\]
must assign value 1 to exactly one projector in every orthonormal basis within the $d$-dimensional subspace. This yields a Kochen-Specker assignment on $K$, contradicting the assumption that $K$ is a Kochen-Specker set.
\end{proof}
\vspace{10pt}

As shown in the proof of Theorem~\ref{thm-01vsKS}, the quantum system $(\mathcal{Q}, \ket{k'})$ is nonclassical even though $\ket{k'}$ is noncontextual. This demonstrates that nonclassicality can be an inherent property of the scenario itself, independent of the contextuality of states.

\begin{figure}[H]
    \centering\includegraphics[width=0.6\linewidth]{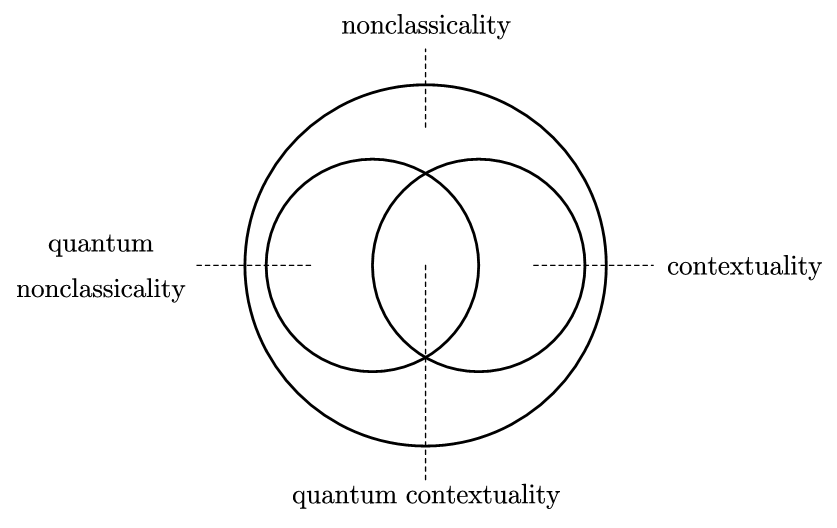}
    \caption{Inclusion relations among four types of nonclassical properties.}
    \label{wenshi}
\end{figure}

Figure~\ref{wenshi} illustrates the inclusion relations among these different types of nonclassical properties.

\subsection{State-independent contextuality}\label{subsec-SIC}

If a quantum scenario $\mathcal{Q}$ can be embedded into a Boolean algebra, then it can be realized by a classical scenario. In such cases, contextual states are required to witness nonclassicality. When all quantum states on $\mathcal{Q}$ are contextual, the scenario exhibits \emph{state-independent contextuality} (SIC).

\vspace{10pt}
\begin{definition}
A quantum scenario $\mathcal{Q}$ is called \textbf{state-independently contextual} if every quantum state $\rho\in s_q(\mathcal{Q})$ is contextual.
\end{definition}
\vspace{10pt}

Note that SIC is a concept specific to quantum states. If all states $p \in s(\mathcal{Q})$ are contextual, then $s_{nc}(\mathcal{Q}) = \emptyset$, which implies that $\mathcal{Q}$ is a Kochen-Specker scenario.

\begin{figure}[H]
    \centering\includegraphics[width=0.55\linewidth]{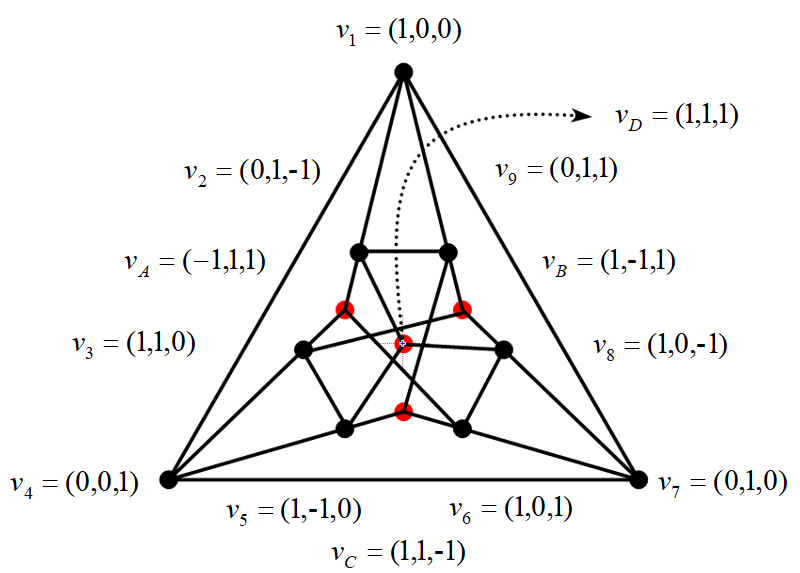}
    \caption{Orthogonality graph of the Yu--Oh set (normalization factors omitted).}
    \label{Yu-Oh}
\end{figure}

A notable example of SIC is provided by the Yu-Oh set \cite{Yu2012State}, which consists of 13 vectors in $\mathbb{C}^3$, as illustrated in Fig.~\ref{Yu-Oh}.

Let $Y = \{\hat{P}_{v_i}\}_{i=1,\dots,9,A,B,C,D}$ be the set of rank-1 projectors corresponding to the Yu-Oh vectors, and let $\mathcal{Q}_{\text{YO}} = \langle Y \rangle$ be the quantum scenario generated by $Y$. It is easily verified that $s_d(\mathcal{Q}_{\text{YO}}) \neq\emptyset$. However, \cite{Yu2012State} showed that
\[
p(\hat{P}_{v_A}) + p(\hat{P}_{v_B}) + p(\hat{P}_{v_C}) + p(\hat{P}_{v_D}) \leq 1 \quad\text{for all } p \in s_{nc}(\mathcal{Q}_{\text{YO}}),
\]
whereas for all quantum states $\rho\in s_q(\mathcal{Q}_{\text{YO}})$,
\[\rho(\hat{P}_{v_A}) + \rho(\hat{P}_{v_B}) + \rho(\hat{P}_{v_C}) + \rho(\hat{P}_{v_D}) = \frac{4}{3},
\]
since $\hat{P}_{v_A} + \hat{P}_{v_B} + \hat{P}_{v_C} + \hat{P}_{v_D} = \frac{4}{3} \mathbf{I}$. Hence, $\mathcal{Q}_{\text{YO}}$ exhibits state-independent contextuality without being a Kochen-Specker scenario.

The projectors in $Y$ serve as both generators and atoms of $\mathcal{Q}_{\text{YO}}$, making $Y$ a perfect generating set. Moreover, if the probabilities on a subset $S \subseteq\mathcal{Q}_{\text{YO}}$ determine the probabilities on all of $Y$, then $S$ is also a perfect generating set. For instance, one can show that the vectors $v_1$, $v_4$, and $v_7$ can be generated by other vectors in Fig.~\ref{Yu-Oh}. Therefore:

\vspace{10pt}
\begin{proposition}
\label{pro-SIC}
The set $S = \{\hat{P}_{v_i}\}_{i \in\{2,3,5,6,8,9,A,B,C,D\}}$ is a perfect generating set of $\mathcal{Q}_{\text{YO}}$. Hence, 10 projectors suffice to witness state-independent contextuality.
\end{proposition}
\begin{proof}
For any $p \in s(\mathcal{Q}_{\text{YO}})$, only need to observe that:
\[
p(\hat{P}_{v_1}) = 1 - p(\hat{P}_{v_2}) - p(\hat{P}_{v_9}), \quad
p(\hat{P}_{v_4}) = 1 - p(\hat{P}_{v_3}) - p(\hat{P}_{v_5}), \quad
p(\hat{P}_{v_7}) = 1 - p(\hat{P}_{v_6}) - p(\hat{P}_{v_8}).
\]
\end{proof}
\vspace{10pt}

It has been shown that any SIC set must contain at least 13 vectors \cite{Cabello2016Quantum}, so Yu-Oh set is the minimal SIC set. Although the minimal quantum scenario exhibiting SIC remains unknown \cite{Budroni2022Kochen}, it can be generated by at most 10 projectors.

\subsection{State-dependent contextuality}\label{subsec-SDC}

If a quantum scenario $\mathcal{Q}$ can be embedded into a Boolean algebra, and only specific quantum states $\rho$ are contextual, then the quantum system $(\mathcal{Q},\rho)$ exhibits \emph{state-dependent contextuality}. Prominent examples include Bell nonlocality \cite{Bell1964On, Clauser1969Proposed} and the Klyachko-Can-Binicioglu-Shumovsky (KCBS) contextuality \cite{Alexander2008Simple}.

\vspace{10pt}
\begin{definition}
A quantum scenario $\mathcal{Q}$ is called \textbf{state-dependently contextual} if there exists a quantum state $\rho\in s_q(\mathcal{Q})$ is contextual.
\end{definition}
\vspace{10pt}

\begin{figure}[H]
    \centering\includegraphics[width=0.27\linewidth]{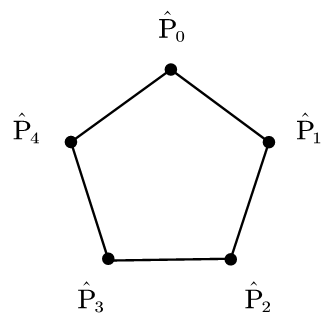}
    \caption{Orthogonality graph of $\{\hat{P}_i \}_{i=0}^4$}
    \label{wujiaoxing}
\end{figure}

KCBS experiment is recognized as the simplest setup witnessing quantum contextuality. It involves measurements of five observables $\{\hat{A}_0, \dots, \hat{A}_4\}$ on a three-dimensional Hilbert space, defined as:
\[\hat{A}_i = 2\hat{S}_{l_i}^2 - \mathbf{I},\]
where $\hat{S}_{l_i}$ is the spin projection operator along direction $l_i$ (with eigenvalues $-1, 0, 1$), and the directions satisfy $l_i \perp l_{i+1}$ (indices modulo 5). The rank-1 projectors $\hat{P}_i := \mathbf{I} - \hat{S}_{l_i}^2$ satisfy $\hat{P}_i \hat{P}_{i+1} = 0$, and the measurement outcomes $\hat{A}_i = -1, 1$ correspond to $\hat{P}_i = 1, 0$, respectively. All observable events in KCBS experiment form a quantum scenario $\mathcal{Q}_{\text{KCBS}} = \langle\{\hat{P}_i \}_{i=0}^4 \rangle$, whose orthogonality graph is shown in Fig.~\ref{wujiaoxing}.

\begin{figure}[H]
     \centering\includegraphics[width=0.5\linewidth]{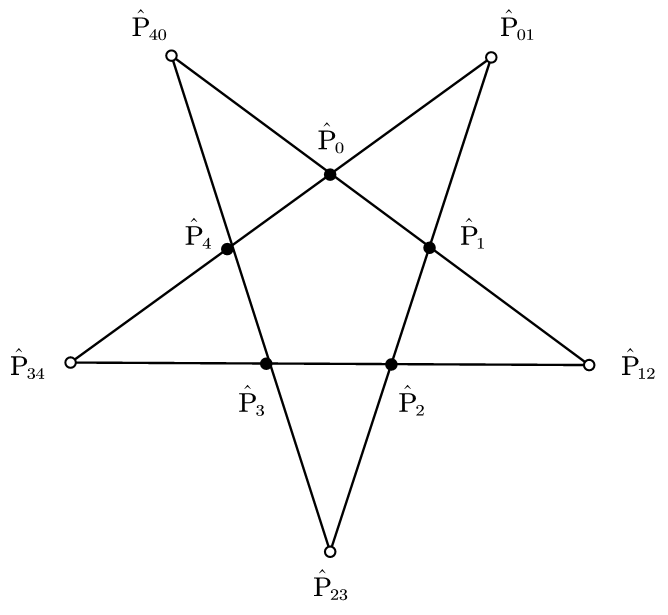}
     \caption{Atom graph $\mathcal{G}_a(\mathcal{Q}_{\text{KCBS}})$}
     \label{KCBS1}
\end{figure}

Using Theorem~\ref{thm:atom_graph_correspondence}, we represent $\mathcal{Q}_{\text{KCBS}}$ via its atom graph. For each $i \in\{0,1,2,3,4\}$, define:
\[\hat{P}_{i,i+1} = \neg(\hat{P}_i \lor\hat{P}_{i+1}),\]
which is a rank-1 projector orthogonal only to $\hat{P}_i$ and $\hat{P}_{i+1}$. The set $\{\hat{P}_i \}_{i=0}^4$ generates exactly five such projectors $\{\hat{P}_{i,i+1} \}_{i=0}^4$. The atom graph $\mathcal{G}_a(\mathcal{Q}_{\text{KCBS}})$ is illustrated in Fig.~\ref{KCBS1}.

Any noncontextual state $p \in s_{nc}(\mathcal{Q}_{\text{KCBS}})$ satisfies the KCBS inequality \cite{Alexander2008Simple, Adan2014Graph}:
\[\sum_{i=0}^4 p(\hat{P}_i) \leq 2.\]
However, quantum mechanics allows violations of this bound. For instance, \cite{Adan2010Contextuality} constructed a quantum state $\rho = \ket{\psi}\!\bra{\psi}$ with $\ket{\psi} = (0,0,1)^T$ and projectors $\hat{P}_i = \ket{v_i}\!\bra{v_i}$, where (omitting normalization):
\begin{align*}
\ket{v_0}&= (1, 0, \sqrt{\cos(\pi/5)})^T, \\
\ket{v_1}&= (\cos(4\pi/5), \sin(4\pi/5), \sqrt{\cos(4\pi/5)})^T, \\
\ket{v_2}&= (\cos(2\pi/5), -\sin(2\pi/5), \sqrt{\cos(\pi/5)})^T, \\
\ket{v_3}&= (\cos(2\pi/5), \sin(2\pi/5), \sqrt{\cos(\pi/5)})^T, \\
\ket{v_4}&= (\cos(4\pi/5), -\sin(4\pi/5), \sqrt{\cos(4\pi/5)})^T.
\end{align*}
A direct computation gives:
\[\sum_{i=0}^4 \rho(\hat{P}_i) = \sqrt{5} > 2,\]
confirming that $\rho$ is contextual.

What is the minimal number of observables required to implement the KCBS experiment, or more generally, to witness quantum contextuality? Within observable-based theories, it is known that 5 observables are needed to witness contextuality in 3-dimensional scenarios, while 4 suffice in general scenarios \cite{Kurzynski2012Entropic, Xu2019Necessary}. However, we note that $\{\hat{P}_{i,i+1}\}_{i=0}^4$ forms a perfect generating set of $\mathcal{Q}_{\text{KCBS}}$, implying that quantum contextuality can be witnessed by only 3 observables.

\vspace{10pt}
\begin{proposition}
\label{pro-KCBS}
The set $\{\hat{P}_{i,i+1}\}_{i=0}^4$ is a perfect generating set of $\mathcal{Q}_{\text{KCBS}}$.
\end{proposition}
\begin{proof}
For any $p \in s(\mathcal{Q}_{\text{KCBS}})$, only need to observe that:
\[
p(\hat{P}_{i,i+1}) = 1 - p(\hat{P}_i) - p(\hat{P}_{i+1}), \quad\text{for } i = 0, 1, 2, 3, 4.
\]
\end{proof}
\vspace{10pt}

\begin{figure}[H]
    \centering\includegraphics[width=0.53\linewidth]{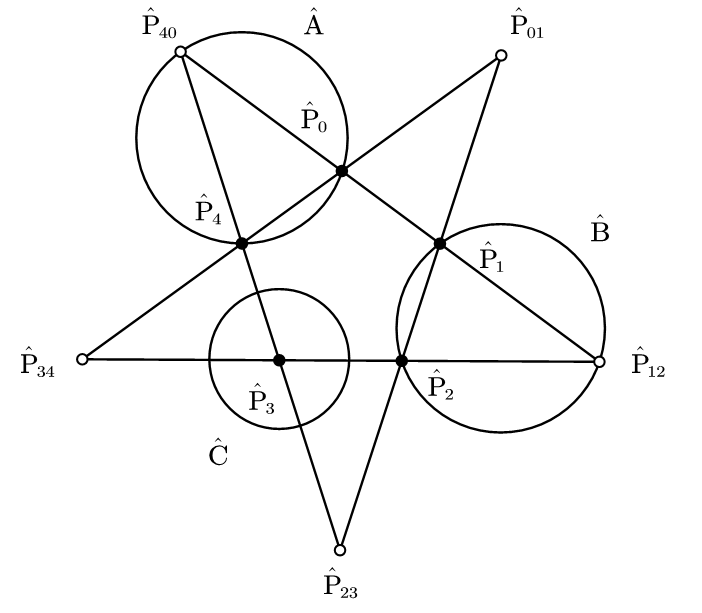}
    \caption{Three observables $\hat{A}, \hat{B}, \hat{C}$ (represented by circumferences) that yield $\{\hat{P}_i\}_{i=0}^4$.}
    \label{KCBS2}
\end{figure}

Define three observables as follows:
\[\hat{A} = a_1 \hat{P}_0 + a_2 \hat{P}_4 + a_3 \hat{P}_{40}, \quad\hat{B} = b_1 \hat{P}_1 + b_2 \hat{P}_2 + b_3 \hat{P}_{12}, \quad\hat{C} = c_1 \hat{P}_3 + c_2 (\neg\hat{P}_3),
\]
where $\{a_i\}, \{b_i\}, \{c_i\}$ are distinct eigenvalues. These observables collectively yield the event set $\{\hat{P}_i\}_{i=0}^4$, as illustrated in Fig.~\ref{KCBS2}.

\vspace{10pt}
\begin{corollary}
\label{cor-3 observables}
Three observables are sufficient to witness quantum contextuality.
\end{corollary}
\vspace{10pt}

Intuitively an quantum experiment which can only measure 2 observables is classical. If the conjecture holds, then Corollary~\ref{cor-3 observables} suggests that 3 is the minimum number of observables required to witness quantum contextuality or nonclassicality.

\section{Conclusion and outlook}\label{sec-conclusion}

In this work, we establish a unified mathematical framework for contextuality and nonclassicality, using the formalism of exclusive partial Boolean algebras, thereby developing the work of connection between quantum logic and contextuality theories. Our framework offers a unified approach to describe and compare general systems $(\mathcal{A},p)$, quantum systems $(\mathcal{Q},\rho)$, and classical systems $(\mathcal{B},p_{\mathcal{B}})$, while preserving the intrinsic logical structure of scenarios. This formulation generalizes existing arguments from observable-based contextuality theories (Theorem~\ref{thm-contextuality}).

We introduced the notion of a \emph{classical embedding} $i_{\mathcal{A}}$, and proved that it provides a unified and minimal classical counterpart $\mathcal{A}^c:=\mathcal{P}(s_d(\mathcal{A}))$ for any finite general scenario $\mathcal{A}$ (Theorems~\ref{thm-imbed} and~\ref{thm-minimal}). Furthermore, we formally established that noncontextual states correspond exactly to convex combinations of deterministic states (Theorem~\ref{thm-convex}). These results yield an economical and precise characterization of nonclassicality (Corollary~\ref{col-classicality}), and demonstrate that contextuality is a sufficient but not necessary condition for nonclassicality (Theorem~\ref{thm-01vsKS}).

Within this framework, we formalized the major categories of quantum contextuality, including Kochen-Specker scenarios, state-independent contextuality (SIC), and state-dependent contextuality. Theoretically, the logical construction adopted here enables the witnessing of contextuality with fewer projectors or observables. Building on well-known examples \cite{Cabello1997Bell,Yu2012State,Alexander2008Simple}, using the concept of \emph{(perfect) generating sets}, we demonstrated that Kochen-Specker scenarios can be generated by 12 projectors (Theorem~\ref{thm-12CEG}), SIC can be witnessed by 10 projectors (Proposition~\ref{pro-SIC}), and quantum contextuality can be witnessed by only 3 observables (Proposition~\ref{pro-KCBS} and Corollary~\ref{cor-3 observables}).

Our framework primarily describes static properties of systems $(\mathcal{A}, p)$. A natural extension would be to incorporate dynamic processes, which can be represented as sequences of state transformations on a fixed scenario:
\[
(\mathcal{A}, p_1) \to (\mathcal{A}, p_2) \to\dots\to (\mathcal{A}, p_n).
\]
For quantum systems $(\mathcal{Q}, \rho)$, the scenario $\mathcal{Q}$ corresponds to fixed experimental devices, while the time evolution is captured by transitions between quantum states. Given the profound relationship between logic and computation, a deeper investigation into the dynamic properties of quantum systems $(\mathcal{Q}, \rho)$ and classical systems $(\mathcal{B}, p_{\mathcal{B}})$ could yield valuable insights for comparing quantum and classical computation.

Furthermore, the logical perspective allows the problem of finding minimal sets of projectors or observables to witness contextuality to be precisely formulated. There exists two unsolvable problems: determining the minimal generating set for a Kochen-Specker scenario, and determining the minimal perfect generating set for a quantum scenario that witnesses contextuality or SIC.

Finally, while the present work focuses on ideal measurements (those directly associated with observables), future research may extend this framework to accommodate non-ideal measurements, such as those described by positive operator-valued measures (POVMs). A key challenge may be the consistent depiction of events (or ``effects") within contexts under general measurements, which may require a substantially different formulation of the logical structure.

\bibliographystyle{unsrturl}
\bibliography{math}

\begin{thebibliography}{10}

\bibitem{Fine1990Einstein}
A.~Fine.
\newblock Einstein and ensembles: Response.
\newblock {\em Foundations of Physics}, 20:967--989, 1990.
\newblock URL: \url{https://doi.org/10.1007/BF00738375}, \href
  {http://dx.doi.org/10.1007/BF00738375} {\path{doi:10.1007/BF00738375}}.

\bibitem{Einstein1935Can}
A.~Einstein, B.~Podolsky, and N.~Rosen.
\newblock Can quantum-mechanical description of physical reality be considered
  complete?
\newblock {\em Phys. Rev.}, 47:777--780, May 1935.
\newblock URL: \url{https://link.aps.org/doi/10.1103/PhysRev.47.777}, \href
  {http://dx.doi.org/10.1103/PhysRev.47.777}
  {\path{doi:10.1103/PhysRev.47.777}}.

\bibitem{Bell1964On}
John~S Bell.
\newblock On the einstein podolsky rosen paradox.
\newblock {\em Physics Physique Fizika}, 1(3):195, 1964.

\bibitem{Clauser1969Proposed}
John~F. Clauser, Michael~A. Horne, Abner Shimony, and Richard~A. Holt.
\newblock Proposed experiment to test local hidden-variable theories.
\newblock {\em Phys. Rev. Lett.}, 23:880--884, Oct 1969.
\newblock URL: \url{https://link.aps.org/doi/10.1103/PhysRevLett.23.880}, \href
  {http://dx.doi.org/10.1103/PhysRevLett.23.880}
  {\path{doi:10.1103/PhysRevLett.23.880}}.

\bibitem{Kochen1967The}
Simon Kochen and E.~P. Specker.
\newblock The problem of hidden variables in quantum mechanics.
\newblock {\em J. Math. Mech.}, 17:59--87, 1967.

\bibitem{Brunner2014Bell}
Nicolas Brunner, Daniel Cavalcanti, Stefano Pironio, Valerio Scarani, and
  Stephanie Wehner.
\newblock Bell nonlocality.
\newblock {\em Rev. Mod. Phys.}, 86:419--478, Apr 2014.
\newblock URL: \url{https://link.aps.org/doi/10.1103/RevModPhys.86.419}, \href
  {http://dx.doi.org/10.1103/RevModPhys.86.419}
  {\path{doi:10.1103/RevModPhys.86.419}}.

\bibitem{Budroni2022Kochen}
Costantino Budroni, Ad\'an Cabello, Otfried G\"uhne, Matthias Kleinmann, and
  Jan-\AA{}ke Larsson.
\newblock Kochen-specker contextuality.
\newblock {\em Rev. Mod. Phys.}, 94:045007, Dec 2022.
\newblock URL: \url{https://link.aps.org/doi/10.1103/RevModPhys.94.045007},
  \href {http://dx.doi.org/10.1103/RevModPhys.94.045007}
  {\path{doi:10.1103/RevModPhys.94.045007}}.

\bibitem{Alexander2008Simple}
Alexander~A. Klyachko, M.~Ali Can, Sinem Binicioglu, and Alexander~S.
  Shumovsky.
\newblock Simple test for hidden variables in spin-1 systems.
\newblock {\em Phys. Rev. Lett.}, 101:020403, Jul 2008.
\newblock URL: \url{https://link.aps.org/doi/10.1103/PhysRevLett.101.020403},
  \href {http://dx.doi.org/10.1103/PhysRevLett.101.020403}
  {\path{doi:10.1103/PhysRevLett.101.020403}}.

\bibitem{Lapkiewicz2011Experimental}
Radek Lapkiewicz, Peizhe Li, Christoph Schaeff, Nathan~K. Langford, Sven
  Ramelow, Marcin Wie\'sniak, and Anton Zeilinger.
\newblock Experimental non-classicality of an indivisible quantum system.
\newblock {\em Nature}, 474:490--493, 2011.
\newblock URL: \url{https://doi.org/10.1038/nature10119}, \href
  {http://dx.doi.org/10.1038/nature10119} {\path{doi:10.1038/nature10119}}.

\bibitem{Kirchmair2009State}
G.~Kirchmair, F.~Z\"ahringer, R.~Gerritsma, M.~Kleinmann, O.~G\"uhne,
  A.~Cabello, R.~Blatt, and C.~F. Roos.
\newblock State-independent experimental test of quantum contextuality.
\newblock {\em Nature}, 460:494--497, 2009.
\newblock URL: \url{https://doi.org/10.1038/nature08172}, \href
  {http://dx.doi.org/10.1038/nature08172} {\path{doi:10.1038/nature08172}}.

\bibitem{Amselem2009State}
Elias Amselem, Magnus R\aa{}dmark, Mohamed Bourennane, and Ad\'an Cabello.
\newblock State-independent quantum contextuality with single photons.
\newblock {\em Phys. Rev. Lett.}, 103:160405, Oct 2009.
\newblock URL: \url{https://link.aps.org/doi/10.1103/PhysRevLett.103.160405},
  \href {http://dx.doi.org/10.1103/PhysRevLett.103.160405}
  {\path{doi:10.1103/PhysRevLett.103.160405}}.

\bibitem{Guhne2010Compatibility}
Otfried G\"uhne, Matthias Kleinmann, Ad\'an Cabello, Jan-\AA{}ke Larsson,
  Gerhard Kirchmair, Florian Z\"ahringer, Rene Gerritsma, and Christian~F.
  Roos.
\newblock Compatibility and noncontextuality for sequential measurements.
\newblock {\em Phys. Rev. A}, 81:022121, Feb 2010.
\newblock URL: \url{https://link.aps.org/doi/10.1103/PhysRevA.81.022121}, \href
  {http://dx.doi.org/10.1103/PhysRevA.81.022121}
  {\path{doi:10.1103/PhysRevA.81.022121}}.

\bibitem{Moussa2010Testing}
Osama Moussa, Colm~A. Ryan, David~G. Cory, and Raymond Laflamme.
\newblock Testing contextuality on quantum ensembles with one clean qubit.
\newblock {\em Phys. Rev. Lett.}, 104:160501, Apr 2010.
\newblock URL: \url{https://link.aps.org/doi/10.1103/PhysRevLett.104.160501},
  \href {http://dx.doi.org/10.1103/PhysRevLett.104.160501}
  {\path{doi:10.1103/PhysRevLett.104.160501}}.

\bibitem{Qu2020Experimental}
Dengke Qu, Pawe\l{} Kurzy\'{n}ski, Dagomir Kaszlikowski, Sadegh Raeisi, Lei
  Xiao, Kunkun Wang, Xiang Zhan, and Peng Xue.
\newblock Experimental entropic test of state-independent contextuality via
  single photons.
\newblock {\em Phys. Rev. A}, 101:060101, Jun 2020.
\newblock URL: \url{https://link.aps.org/doi/10.1103/PhysRevA.101.060101},
  \href {http://dx.doi.org/10.1103/PhysRevA.101.060101}
  {\path{doi:10.1103/PhysRevA.101.060101}}.

\bibitem{Mark2014Contextuality}
Mark Howard, Joel Wallman, Victor Veitch, and Joseph Emerson.
\newblock Contextuality supplies the 'magic' for quantum computation.
\newblock {\em Nature}, 510:351--355, 2014.
\newblock URL: \url{https://doi.org/10.1038/nature13460}, \href
  {http://dx.doi.org/10.1038/nature13460} {\path{doi:10.1038/nature13460}}.

\bibitem{Gupta2023Quantum}
Shashank Gupta, Debashis Saha, Zhen-Peng Xu, Ad\'an Cabello, and A.~S.
  Majumdar.
\newblock Quantum contextuality provides communication complexity advantage.
\newblock {\em Phys. Rev. Lett.}, 130:080802, Feb 2023.
\newblock URL: \url{https://link.aps.org/doi/10.1103/PhysRevLett.130.080802},
  \href {http://dx.doi.org/10.1103/PhysRevLett.130.080802}
  {\path{doi:10.1103/PhysRevLett.130.080802}}.

\bibitem{Pirandola2020Advances}
S.~Pirandola, U.~L. Andersen, L.~Banchi, M.~Berta, D.~Bunandar, R.~Colbeck,
  D.~Englund, T.~Gehring, C.~Lupo, C.~Ottaviani, J.~L. Pereira, M.~Razavi,
  J.~Shamsul Shaari, M.~Tomamichel, V.~C. Usenko, G.~Vallone, P.~Villoresi, and
  P.~Wallden.
\newblock Advances in quantum cryptography.
\newblock {\em Adv. Opt. Photon.}, 12(4):1012--1236, Dec 2020.
\newblock URL: \url{https://opg.optica.org/aop/abstract.cfm?URI=aop-12-4-1012},
  \href {http://dx.doi.org/10.1364/AOP.361502} {\path{doi:10.1364/AOP.361502}}.

\bibitem{Kurzynski2012Entropic}
P.~Kurzy\'nski, R.~Ramanathan, and D.~Kaszlikowski.
\newblock Entropic test of quantum contextuality.
\newblock {\em Phys. Rev. Lett.}, 109:020404, Jul 2012.
\newblock URL: \url{https://link.aps.org/doi/10.1103/PhysRevLett.109.020404},
  \href {http://dx.doi.org/10.1103/PhysRevLett.109.020404}
  {\path{doi:10.1103/PhysRevLett.109.020404}}.

\bibitem{Fritz2013Entropic}
Tobias Fritz and Rafael Chaves.
\newblock Entropic inequalities and marginal problems.
\newblock {\em IEEE Transactions on Information Theory}, 59(2):803--817, 2013.
\newblock \href {http://dx.doi.org/10.1109/TIT.2012.2222863}
  {\path{doi:10.1109/TIT.2012.2222863}}.

\bibitem{Abramsky2011sheaf}
Samson Abramsky and Adam Brandenburger.
\newblock The sheaf-theoretic structure of non-locality and contextuality.
\newblock {\em New Journal of Physics}, 13(11):113036, nov 2011.
\newblock URL: \url{https://dx.doi.org/10.1088/1367-2630/13/11/113036}, \href
  {http://dx.doi.org/10.1088/1367-2630/13/11/113036}
  {\path{doi:10.1088/1367-2630/13/11/113036}}.

\bibitem{Adan2014Graph}
Ad\'an Cabello, Simone Severini, and Andreas Winter.
\newblock Graph-theoretic approach to quantum correlations.
\newblock {\em Phys. Rev. Lett.}, 112:040401, Jan 2014.
\newblock URL: \url{https://link.aps.org/doi/10.1103/PhysRevLett.112.040401},
  \href {http://dx.doi.org/10.1103/PhysRevLett.112.040401}
  {\path{doi:10.1103/PhysRevLett.112.040401}}.

\bibitem{Acin2015A}
Antonio Acin, Tobias Fritz, Anthony Leverrier, and Ana~Bel\'en Sainz.
\newblock A combinatorial approach to nonlocality and contextuality.
\newblock {\em Communications in Mathematical Physics}, 334:533--628, 2015.
\newblock URL: \url{https://doi.org/10.1007/s00220-014-2260-1}, \href
  {http://dx.doi.org/10.1007/s00220-014-2260-1}
  {\path{doi:10.1007/s00220-014-2260-1}}.

\bibitem{Amaral2018On}
Barbara Amaral and Marcelo Terra~Cunha.
\newblock {\em On graph approaches to contextuality and their role in quantum
  theory}.
\newblock SpringerBriefs in Mathematics. Springer, Cham, 2018.
\newblock URL: \url{https://doi.org/10.1007/978-3-319-93827-1}, \href
  {http://dx.doi.org/10.1007/978-3-319-93827-1}
  {\path{doi:10.1007/978-3-319-93827-1}}.

\bibitem{Abramsky2020The}
Samson Abramsky and Rui~Soares Barbosa.
\newblock The logic of contextuality.
\newblock In Christel Baier and Jean Goubault-Larrecq, editors, {\em 29th EACSL
  Annual Conference on Computer Science Logic (CSL 2021)}, volume 183 of {\em
  Leibniz International Proceedings in Informatics (LIPIcs)}, pages 5:1--5:18,
  Dagstuhl, Germany, 2021. Schloss Dagstuhl -- Leibniz-Zentrum f{\"u}r
  Informatik.
\newblock URL:
  \url{https://drops.dagstuhl.de/entities/document/10.4230/LIPIcs.CSL.2021.5},
  \href {http://dx.doi.org/10.4230/LIPIcs.CSL.2021.5}
  {\path{doi:10.4230/LIPIcs.CSL.2021.5}}.

\bibitem{Liu2025Atom}
Songyi Liu, Yongjun Wang, Baoshan Wang, Jian Yan, and Heng Zhou.
\newblock Atom graph, partial boolean algebra and quantum contextuality.
\newblock {\em Quantum Information Processing}, 24:12, 2025.

\bibitem{Yu2012State}
Sixia Yu and C.~H. Oh.
\newblock State-independent proof of kochen-specker theorem with 13 rays.
\newblock {\em Phys. Rev. Lett.}, 108:030402, Jan 2012.
\newblock URL: \url{https://link.aps.org/doi/10.1103/PhysRevLett.108.030402},
  \href {http://dx.doi.org/10.1103/PhysRevLett.108.030402}
  {\path{doi:10.1103/PhysRevLett.108.030402}}.

\bibitem{Cabello2016Quantum}
Ad\'an Cabello, Matthias Kleinmann, and Jos\'e~R Portillo.
\newblock Quantum state-independent contextuality requires 13 rays.
\newblock {\em Journal of Physics A: Mathematical and Theoretical},
  49(38):38LT01, aug 2016.
\newblock URL: \url{https://doi.org/10.1088/1751-8113/49/38/38LT01}, \href
  {http://dx.doi.org/10.1088/1751-8113/49/38/38LT01}
  {\path{doi:10.1088/1751-8113/49/38/38LT01}}.

\bibitem{Birkhoff1936The}
Garrett Birkhoff and John~Von Neumann.
\newblock The logic of quantum mechanics.
\newblock {\em Annals of Mathematics}, 37(4):823--843, 1936.
\newblock URL: \url{http://www.jstor.org/stable/1968621}.

\bibitem{Coecke2002Quantum}
Bob Coecke.
\newblock Quantum logic in intuitionistic perspective.
\newblock {\em Studia Logica}, 70:411--440, 2002.
\newblock URL: \url{https://doi.org/10.1023/A:1015106515413}, \href
  {http://dx.doi.org/10.1023/A:1015106515413}
  {\path{doi:10.1023/A:1015106515413}}.

\bibitem{Isham1998Topos}
C.~J. Isham and J.~Butterfield.
\newblock Topos perspective on the kochen-specker theorem: I. quantum states as
  generalized valuations.
\newblock {\em International Journal of Theoretical Physics}, 37:2669--2733,
  1998.
\newblock URL: \url{https://doi.org/10.1023/A:1026680806775}, \href
  {http://dx.doi.org/10.1023/A:1026680806775}
  {\path{doi:10.1023/A:1026680806775}}.

\bibitem{Doering2010Topos}
Andreas Doering.
\newblock Topos quantum logic and mixed states.
\newblock {\em Electronic Notes in Theoretical Computer Science},
  270(2):59--77, 2011.
\newblock Proceedings of the 6th International Workshop on Quantum Physics and
  Logic (QPL 2009).
\newblock URL:
  \url{https://www.sciencedirect.com/science/article/pii/S1571066111000247},
  \href {http://dx.doi.org/https://doi.org/10.1016/j.entcs.2011.01.023}
  {\path{doi:https://doi.org/10.1016/j.entcs.2011.01.023}}.

\bibitem{Frembs2023Gleason}
Markus Frembs and Andreas D\"{o}ring.
\newblock Gleason’s theorem for composite systems.
\newblock {\em Journal of Physics A: Mathematical and Theoretical},
  56(44):445303, oct 2023.
\newblock URL: \url{https://dx.doi.org/10.1088/1751-8121/acfbcb}, \href
  {http://dx.doi.org/10.1088/1751-8121/acfbcb}
  {\path{doi:10.1088/1751-8121/acfbcb}}.

\bibitem{Frembs2025Coming}
Markus Frembs.
\newblock Coming full circle -- a unified framework for kochen-specker
  contextuality, 2025.
\newblock URL: \url{https://arxiv.org/abs/2501.09750}, \href
  {http://arxiv.org/abs/2501.09750} {\path{arXiv:2501.09750}}.

\bibitem{Foulis2006Quantum}
D.J. Foulis, R.J. Greechie, M.~Louisa Dalla~Chiara, and R.~Giuntini.
\newblock {\em Quantum Logic.}
\newblock University of Massachusetts, 2006.
\newblock URL:
  \url{https://search.ebscohost.com/login.aspx?direct=true&amp;db=edselc&amp;AN=edselc.2-52.0-84889442066&amp;lang=zh-cn&amp;site=eds-live}.

\bibitem{Kochen2015Reconstruction}
Simon Kochen.
\newblock A reconstruction of quantum mechanics.
\newblock {\em Foundations of Physics}, 45:557--590, 10 1994.
\newblock URL: \url{https://doi.org/10.1007/s10701-015-9886-5}, \href
  {http://dx.doi.org/10.1007/s10701-015-9886-5}
  {\path{doi:10.1007/s10701-015-9886-5}}.

\bibitem{Cabello1997Bell}
Ad\'an Cabello, Jos\'eM. Estebaranz, and Guillermo Garc\'ia-Alcaine.
\newblock Bell-{K}ochen-{S}pecker theorem: A proof with 18 vectors.
\newblock {\em Physics Letters A}, 212(4):183--187, 1996.
\newblock URL:
  \url{https://www.sciencedirect.com/science/article/pii/037596019600134X},
  \href {http://dx.doi.org/https://doi.org/10.1016/0375-9601(96)00134-X}
  {\path{doi:https://doi.org/10.1016/0375-9601(96)00134-X}}.

\bibitem{Xu2020Proof}
Zhen-Peng Xu, Jing-Ling Chen, and Otfried G\"uhne.
\newblock Proof of the peres conjecture for contextuality.
\newblock {\em Phys. Rev. Lett.}, 124:230401, Jun 2020.
\newblock URL: \url{https://link.aps.org/doi/10.1103/PhysRevLett.124.230401},
  \href {http://dx.doi.org/10.1103/PhysRevLett.124.230401}
  {\path{doi:10.1103/PhysRevLett.124.230401}}.

\bibitem{Xu2019Necessary}
Zhen-Peng Xu and Ad\'an Cabello.
\newblock Necessary and sufficient condition for contextuality from
  incompatibility.
\newblock {\em Phys. Rev. A}, 99:020103, Feb 2019.
\newblock URL: \url{https://link.aps.org/doi/10.1103/PhysRevA.99.020103}, \href
  {http://dx.doi.org/10.1103/PhysRevA.99.020103}
  {\path{doi:10.1103/PhysRevA.99.020103}}.

\bibitem{Van2012Noncommutativity}
Benno van~den Berg and Chris Heunen.
\newblock Noncommutativity as a colimit.
\newblock {\em Applied Categorical Structures}, 20:393--414, 2012.
\newblock URL: \url{https://doi.org/10.1007/s10485-011-9246-3}, \href
  {http://dx.doi.org/10.1007/s10485-011-9246-3}
  {\path{doi:10.1007/s10485-011-9246-3}}.

\bibitem{Abramsky2015Contextuality}
Samson Abramsky, Rui Soares~Barbosa, Kohei Kishida, Raymond Lal, and Shane
  Mansfield.
\newblock Contextuality, cohomology and paradox.
\newblock In Stephan Kreutzer, editor, {\em 24th EACSL Annual Conference on
  Computer Science Logic (CSL 2015)}, volume~41 of {\em Leibniz International
  Proceedings in Informatics (LIPIcs)}, pages 211--228, Dagstuhl, Germany,
  2015. Schloss Dagstuhl -- Leibniz-Zentrum f{\"u}r Informatik.
\newblock URL:
  \url{https://drops.dagstuhl.de/entities/document/10.4230/LIPIcs.CSL.2015.211},
  \href {http://dx.doi.org/10.4230/LIPIcs.CSL.2015.211}
  {\path{doi:10.4230/LIPIcs.CSL.2015.211}}.

\bibitem{Ramanathan2012Generalized}
Ravishankar Ramanathan, Akihito Soeda, Pawe\l Kurzy\'{n}ski, and Dagomir
  Kaszlikowski.
\newblock Generalized monogamy of contextual inequalities from the
  no-disturbance principle.
\newblock {\em Phys. Rev. Lett.}, 109:050404, Aug 2012.
\newblock URL: \url{https://link.aps.org/doi/10.1103/PhysRevLett.109.050404},
  \href {http://dx.doi.org/10.1103/PhysRevLett.109.050404}
  {\path{doi:10.1103/PhysRevLett.109.050404}}.

\bibitem{Popescu1994Quantum}
S.~Popescu and D.~Rohrlich.
\newblock Quantum nonlocality as an axiom.
\newblock {\em Foundations of Physic}, 24:379--385, 1994.
\newblock URL: \url{https://doi.org/10.1007/BF02058098}.

\bibitem{Adan2012Specker}
Ad/'an Cabello.
\newblock Specker's fundamental principle of quantum mechanics, 2012.
\newblock URL: \url{https://arxiv.org/abs/1212.1756}, \href
  {http://arxiv.org/abs/1212.1756} {\path{arXiv:1212.1756}}.

\bibitem{Fritz2013Local}
T.~Fritz, A.B. Sainz, R.~Augusiak, J~Bohr Brask, R.~Chaves, A.~Leverrier, and
  A.~Acín.
\newblock Local orthogonality as a multipartite principle for quantum
  correlations.
\newblock {\em Nature Communications}, 4:2263, 2013.
\newblock URL: \url{https://doi.org/10.1038/ncomms3263}, \href
  {http://dx.doi.org/10.1038/ncomms3263} {\path{doi:10.1038/ncomms3263}}.

\bibitem{Gleason1957Measures}
Andrew Gleason.
\newblock Measures on the closed subspaces of a hilbert space.
\newblock {\em Indiana Univ. Math. J.}, 6:885--893, 1957.

\bibitem{Lisonek2014Kochen}
Petr Lison\v{e}k, Piotr Badzia\c{}g, Jos\'e~R. Portillo, and Ad\'an Cabello.
\newblock Kochen-specker set with seven contexts.
\newblock {\em Phys. Rev. A}, 89:042101, Apr 2014.
\newblock URL: \url{https://link.aps.org/doi/10.1103/PhysRevA.89.042101}, \href
  {http://dx.doi.org/10.1103/PhysRevA.89.042101}
  {\path{doi:10.1103/PhysRevA.89.042101}}.

\bibitem{Li2022An}
Zhengyu Li, Curtis Bright, and Vijay Ganesh.
\newblock An sc-square approach to the minimum kochen–specker problem.
\newblock In {\em The 7th International Workshop on Satisfiability Checking and
  Symbolic Computation}, August 12 2022.

\bibitem{Cabello2015Necessary}
Ad\'an Cabello, Matthias Kleinmann, and Costantino Budroni.
\newblock Necessary and sufficient condition for quantum state-independent
  contextuality.
\newblock {\em Phys. Rev. Lett.}, 114:250402, Jun 2015.
\newblock URL: \url{https://link.aps.org/doi/10.1103/PhysRevLett.114.250402},
  \href {http://dx.doi.org/10.1103/PhysRevLett.114.250402}
  {\path{doi:10.1103/PhysRevLett.114.250402}}.

\end{thebibliography}

\appendix
\setcounter{definition}{0}
\renewcommand{\thedefinition}{A\arabic{definition}}
\setcounter{theorem}{0}
\renewcommand{\thetheorem}{A\arabic{theorem}}

\section{Sheaf-theoretic approach of contextuality}\label{appendix-A}

We recall the basic concepts from the sheaf-theoretic approach to contextuality here \cite{Abramsky2011sheaf, Abramsky2020The}.

\vspace{10pt}
\begin{definition}
A \textbf{graphical measurement scenario} is a triple $(X, \frown, O)$, where:
\begin{enumerate}
  \item$X$ is a set of measurements,
  \item$\frown$ is a reflexive and symmetric compatibility relation on $X$,
  \item$O = (O_x)_{x \in X}$ is a family of finite outcome sets for each measurement.
\end{enumerate}
A \textbf{context} is a subset $C \subseteq X$ of pairwise compatible measurements, i.e., a clique under $\frown$. The set of all contexts is denoted by $Kl(\frown)$.
\end{definition}

\vspace{10pt}
\begin{definition}
Let $(X, \frown, O)$ be a measurement scenario. A \textbf{no-signaling empirical model} is a family $(p_C)_{C \in Kl(\frown)}$ of probability distributions, where each $p_C$ is defined on the set of joint outcomes $\mathcal{E}(C) := \prod_{x \in C} O_x$ and satisfies the following local consistency condition: for any contexts $C, C' \in Kl(\frown)$ with $C \subseteq C'$,
\[p_{C'}|_C(s) = \sum_{\substack{t \in\mathcal{E}(C') \\ t|_C = s}} p_{C'}(t)=p_C(s) \quad\text{for all }s \in\mathcal{E}(C),\]
where $p_{C'}|_C$ denotes the marginal distribution of $p_{C'}$ on $C$.
An empirical model is \textbf{noncontextual} if there exists a global probability distribution $d$ on $\mathcal{E}(X) := \prod_{x \in X} O_x$ such that $d|_C = p_C$ for every context $C \in Kl(\frown)$.
\end{definition}
\vspace{10pt}

\begin{theorem}[\cite{Abramsky2011sheaf}]
\label{thm-sheaf-nc}
A no-signaling empirical model $p$ is noncontextual if and only if it admits a NCHV model.
\end{theorem}
\vspace{10pt}

\end{document}